%% file: acl_latex.tex
\documentclass[11pt]{article}
\usepackage[final]{acl}

\usepackage{times}
\usepackage{latexsym}
\usepackage[T1]{fontenc}
\usepackage[utf8]{inputenc}
\usepackage{microtype}
\usepackage{inconsolata}
\usepackage{graphicx}
\usepackage{longtable}
\usepackage{booktabs}
\usepackage{amssymb}
\usepackage{amsmath}
\usepackage{enumitem}
\usepackage{multirow}
\usepackage{tcolorbox}
\usepackage{placeins}
\usepackage{natbib}
\setkeys{Gin}{draft=false}

\title{Talking to Itself While Coding: What Makes Comments Help Code Generation?}

\author{
  \textbf{Dangfeng Pan\textsuperscript{1}},
  \textbf{Zhensu Sun\textsuperscript{2}}\thanks{Corresponding author.},
  \textbf{Cenyuan Zhang\textsuperscript{1}},
  \textbf{David Lo\textsuperscript{2}},
  \textbf{Xiaoning Du\textsuperscript{1}}
\\
  \textsuperscript{1}Monash University, Australia,
  \textsuperscript{2}Singapore Management University, Singapore
}

\begin{document}
\maketitle

\input{sections/00_abstract}

\input{sections/01_introduction}
\input{sections/02_related}
\input{sections/03_observation}
\input{sections/04_intervention}
\input{sections/05_conclusion}

\input{sections/limitations}
\input{sections/ethics}
\input{sections/acknowledgments}

\bibliography{refs}

\appendix
\input{appendix/A_stats}

\input{appendix/B_prompts}

\input{appendix/C_models_compute}

\input{appendix/D_supplementary}

\end{document}

%% file: sections/00_abstract.tex
\begin{abstract}
Large Language Models (LLMs) often generate natural-language comments while writing code, and these comments become part of the context used to generate the code that follows.
However, it remains unclear which properties of comments affect code-generation performance.
We study this question through observational analyses and controlled interventions.
On LiveCodeBench, neither comment frequency nor broad comment intent reliably predicts pass@1.
We then prefill weaker recipient models with comment blocks written by stronger source models, allowing us to separate comment surface form from the solution content they convey.
Comments from source solutions that pass the tests raise recipient pass@1 by 17.2\% on average.
In contrast, comments describing failed solutions provide no reliable gain, while comments written for a different problem reduce pass@1 by 20.8\%.
Finally, across a wide range of models and prompt variants, most recipient models show no significant recovery of the external-comment gain, and the best case recovers only 24\%.
These results show that comments help code generation not merely because they are comments, but because they can provide correct solution content that prompting cannot reliably elicit.
\end{abstract}

%% file: sections/01_introduction.tex
\section{Introduction}
\label{sec:intro}

Code comments have long been studied as text for human readers. They explain rationale to future maintainers and share intent with collaborators.
Empirical work has shown that informative comments measurably improve human program comprehension~\citep{tenny1988}.
Subsequent taxonomies classify comments by intent and trace how each category supports a distinct maintenance activity~\citep{storey2005,pascarella2017}.

In LLM-based code generation, the reader changes from a human to the model itself.
The comments and the code are emitted in the same forward pass by the same model.
The model reads its own comments as context for the code that follows.
We call these self-emitted comments, and their role in the model's own solution is not obvious.
These comments are not free: they are billed as output tokens \citep{pan2025hiddencostreadabilitycode}, and whether they can be safely suppressed depends on whether they causally help.

Do these comments causally affect the model's own pass@1?
Prior work has shown that natural-language preambles in code generation can help \citep{jiang2024selfplanning, jiang2024mango, shi2024outlines, di2025bidirectional, jiang2026thinkanywhere}, but has not isolated which of their properties—surface form, line order, topic, intent, or the correctness of the solution they describe—is responsible.
As a first observational check, we ask whether the most visible properties of natural comments provide a simple explanation for pass@1.
We begin with comment volume and surface form.
If these properties provided such an explanation, we would expect some consistent association between commenting behavior and performance.
On LiveCodeBench~\citep{jain2024lcb}, OpenReasoning-Nemotron-32B and Gemini-2.5-Pro-03-25 reach 81.0\% and 81.5\% pass@1 respectively, while self-emitted comments account for 0.1\% of Nemotron's output tokens and 71.4\% of Gemini's: a 700-fold gap that is invisible in pass@1.
Among the 16 models scoring above 80.0\% pass@1, commenting rates range from under 2.0\% (4 models) to over 30.0\% (5 models), showing no apparent link to pass@1.

In addition to the volume of comments, the intent of the comments also fails to track pass@1.
Specifically, we label every comment across 13 models along the WHY/WHAT axis from comment taxonomies~\citep{storey2005,pascarella2017}—whether it explains a choice or describes what the code does, and check whether models that explain choices more often tend to score higher.
We find that stronger models do explain choices a bit more often ($\rho = +0.61$, raw $p = 0.027$), but the trend does not survive Bonferroni correction.
Therefore, from the natural distribution alone, neither comment volume/form nor coarse intent provides a reliable explanation for pass@1.
This observational result cannot determine whether comments causally matter, so we turn to a controlled intervention (Section~\ref{sec:intervention}).
We prefill a weaker \emph{recipient} model's decoding context with a comment block from a stronger \emph{source} model, and let the recipient continue from there, then vary only the comment's content while holding the recipient, prompt, and problem fixed.
The source is chosen to be strong for a reason: its comments are correct when its own solution passes the tests and wrong when it fails, letting us control content correctness directly instead of merely measuring it.
The external prefix is thus a measurement instrument rather than a model of natural self-commenting: it identifies what property a comment must have to help, and Section~\ref{sec:self-elicit} then asks whether the model's own comments can supply that property.

The intervention separates which property of the comment moves pass@1.
Correct comments raise the recipient's pass@1 by $17.2\%$ on average across the 12 (recipient, source) pairs.
Same-form comments with wrong content leave pass@1 unchanged.
Comment-shaped text on the wrong topic damages pass@1 by $20.8\%$, and length-matched random text damages it by $17.9\%$.
Pass@1 changes with whether the comment describes a correct solution, not with what the comment looks like.

A natural follow-up question is whether the recipient can produce a comparable comment from its own prompt, without an external source.
Across 14 models and 10 prompt variants (Section~\ref{sec:self-elicit}), the answer is largely no.
Models follow what the prompt asks at the surface, such as producing more or fewer comments, or adopting different styles.
However, the best prompt recovers at most about $24\%$ of the external-comment lift, and the same instruction can move pass@1 in opposite directions on different recipients.
Prompt engineering controls what the comments look like, but cannot reliably control whether their content describes a correct solution.

\begin{figure*}[t]
\centering
\makebox[\linewidth][c]{%
  \includegraphics[
    width=1.08\linewidth,
    trim=18 45 18 48,
    clip
  ]{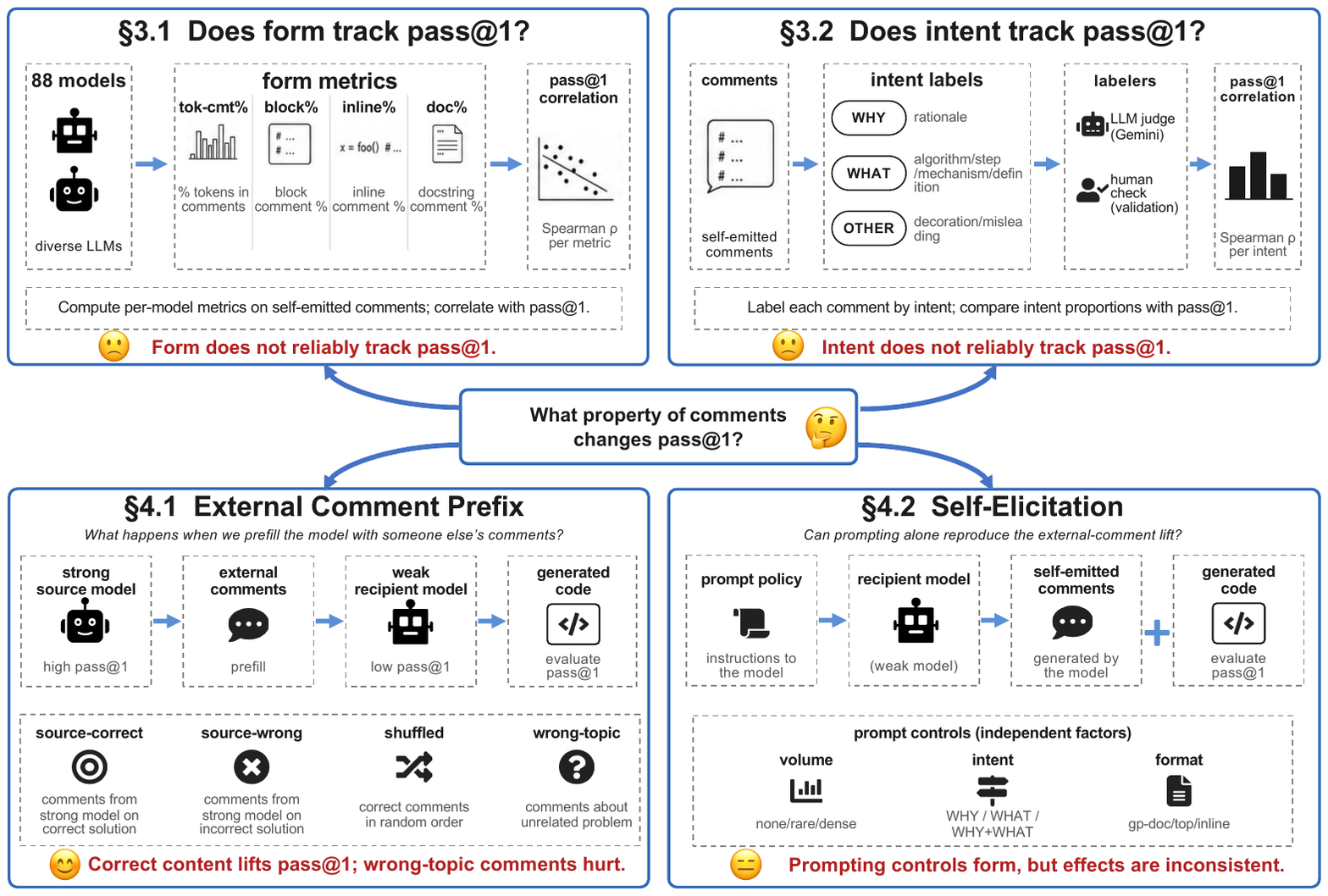}
}
\caption{The four core sub-experiments. Section~\ref{sec:form} profiles comment form across 88 models; Section~\ref{sec:intent} profiles comment intent across 13 models. Section~\ref{sec:decomposition} attaches a stronger source's comment block to a weaker recipient and varies its content. Section~\ref{sec:self-elicit} asks whether self-prompting can elicit the same lift.}
\label{fig:study-overview}
\end{figure*}

The overall structure of our study is depicted in Figure~\ref{fig:study-overview}.
In summary, the natural distribution of self-emitted comments does not track pass@1.
Under controlled intervention, a correct comment from a stronger model lifts a weaker model's pass@1 while prompting the weaker model to write its own comments falls short.
Talking to itself is not enough, and what matters is whether the talk carries correct solution content.
To facilitate reproducibility, we release
our code, generation outputs, per-problem pass/fail evaluations, and per-cell statistical summaries at \url{https://github.com/pdfCN/What-Makes-Comments-Help-Code-Generation}.

%% file: sections/02_related.tex
\section{Related Work on Code Comments in Code Generation}
\label{sec:related}
Code comments have historically been studied as text written for human readers \citep{tenny1988, storey2005, pascarella2017}.
However, in LLM-based code generation, comments can also become context for the model itself.
Existing work has explored two related directions.
In the first, the comment is input read by the model.
Rationale comments help auto bug-fix \citep{vitale2026impactcodecommentsautomated}, removing comments lifts some downstream tasks \citep{imani2025context, liu2024repoqa}, mismatched comments degrade unit-test generation \citep{macke-doyle-2024-testing}, and probes find LLM-emitted comments themselves often inaccurate \citep{kang2024inaccurate}.
In the second direction, the model itself emits natural language as a preamble to its own code.
Plan-then-code prompting \citep{jiang2024selfplanning}, comment-then-code training \citep{jiang2024mango}, natural-language outlines above code \citep{shi2024outlines}, bidirectional comment-code grounding \citep{di2025bidirectional}, and placement variation during decoding \citep{jiang2026thinkanywhere} all show that getting the model to produce natural language before code helps.

None of these works isolates which property of the preamble is responsible for the effect.
Candidates include the surface form, the line order, the topic, the intent and the correctness of the described solution content.
Whether any property is dominant remains open.

We adapt the approach from In-Context Learning (ICL) studies that pull demonstrations apart along format, label distribution, order, and
surface structure \citep{min2022rethinking, lu2022ordered, sclar2024quantifying, li2025structurenotcontent}.
\citet{min2022rethinking} found that ground-truth labels matter little in ICL demonstrations, while a counterpoint shows that labels can still matter when demonstrations are chosen carefully \citep{yoo2022groundtruth}.
Our intervention attaches a comment block from a stronger source, varies one property at a time while holding the others fixed, and identifies content correctness as the dominant property in the comment interface.
In this decomposition our finding lands closer to \citet{yoo2022groundtruth}: it can be read as a reproduction and extension of the ICL content-versus-form question through the comment interface.

%% file: sections/03_observation.tex
\section{Observing Self-Emitted Comments}
\label{sec:observation}

We first ask whether the natural distribution of self-emitted comments shows any pattern with pass@1.
If commenting style differed sharply between strong and weak models, we would already have direct evidence that comments support the model's own generation.
If not, the question demands a controlled experiment.
We test this on two axes.
The first is \textbf{form}, the surface shape of the comments (Section~\ref{sec:form}).
The second is \textbf{intent}, what each comment line says (Section~\ref{sec:intent}).

\subsection{Does form track pass@1?}
\label{sec:form}

\label{sec:form-setup}
We analyse 88 models on LiveCodeBench v6 \citep{jain2024lcb}, listed in Appendix~\ref{app:model-pool}.
Of these, 14 are run by ourselves and 74 come from the public leaderboard \citep{lcb_submissions}.
We run 10 via OpenRouter and 4 locally with vLLM.
We measure commenting style with four metrics computed under a single shared tokenizer: \textbf{tok-cmt\%} (fraction of output tokens inside a comment) and the PEP-8 partition \citep{pep8} into \textbf{block\%}, \textbf{inline\%}, and \textbf{doc\%} (formal definitions in Appendix~\ref{app:form-metrics}).

We test whether form tracks pass@1 at two layers.
Across models, Table~\ref{tab:toptier-spread} (Appendix~\ref{app:full-form-profile}) reports the spread of each form axis at matched pass@1.
Within each model, we compute the Spearman rank correlation $\rho$ \citep{spearman1904} between each per-problem form feature and pass@1.

\paragraph{At matched pass@1, commenting volume and form vary widely.}
\label{sec:form-no-comments}
Of the 16 top-tier models with pass@1 $\geq 80.0\%$, 4 emit fewer than $2.0\%$ comment tokens while 5 exceed $30.0\%$: \texttt{tok-cmt\%} spans $0.1\%$ (OpenReasoning-Nemotron-32B) to $71.4\%$ (Gemini-2.5-Pro-03-25), a 700-fold spread, and the other three axes vary just as widely (Table~\ref{tab:toptier-spread}).
\label{sec:form-l1}
The same pattern extends across all 88 models: $10.2\%$ emit fewer than $2.0\%$ comment tokens and $45.5\%$ emit more than $20.0\%$.
Low comment volume is not by itself a signal of model weakness, and cross-model comparison alone cannot tell us whether commenting tracks pass@1.

\paragraph{A model's commenting habits look the same on problems it solves and problems it fails.}
\label{sec:form-l2}

On the three format axes \texttt{block\%}, \texttt{inline\%}, and \texttt{doc\%}, the strongest per-model $|\rho|$ across the 88-model pool is only $0.32$ (Appendix~\ref{app:within-model-rho}), and at most $13.0\%$ of models exceed $|\rho|{=}0.20$, the conventional small-effect threshold.
The \texttt{tok-cmt\%} axis shows a slightly larger upper tail, with $22.0\%$ of models exceeding $|\rho|{=}0.20$ and a maximum of $0.53$, still too small to change the conclusion.
Form does not track pass@1 at either layer.

\subsection{Does intent track pass@1?}
\label{sec:intent}

The form of a comment does not capture what it says.
Two block comments can look identical on every form axis but mean very different things.
For example, \texttt{\# Use dynamic programming because subproblems repeat} explains the reasoning, while \texttt{\# Sort the array} only restates what the next line of code does.
We therefore ask whether intent tracks pass@1.

We label each comment line along the WHY/WHAT axis standard in code-comment taxonomies \citep{storey2005,pascarella2017}, plus an OTHER catch-all (about $10\%$ of lines); the three buckets aggregate a finer seven-class codebook whose per-label reliability guides the bucket boundaries (codebook and mapping in Appendix~\ref{app:kappa}).
We compute intent fractions on a 13-model pool spanning the LCB pass@1 spectrum, with 8 self-run models and 5 leaderboard submissions (Appendix~\ref{app:model-pool}).
Labels come from an LLM judge, \texttt{Gemini-3.1-Flash-Lite}, validated against a stratified 100-snippet human-labelled calibration set (judge prompt in Appendix~\ref{app:prompt-judge}): the anchor labels of WHY and WHAT reach Cohen's $\kappa$ \citep{cohen1960} of $+0.74$ and $+0.72$, and bucket-level agreement is in Appendix~\ref{app:kappa}.
For each intent, we compute the cross-model Spearman~$\rho$ between the model's mean intent fraction and its pass@1.
The mean is taken over samples that contain at least one comment, to remove the volume confound.
We apply Bonferroni correction over the three intents.

\paragraph{Intent does not track pass@1 either.}
\label{sec:intent-result}

The rank correlations are positive on WHY ($\rho={+}0.61$) and negative on WHAT ($\rho={-}0.20$): stronger models in our pool emit more rationale and less procedural description than weaker ones.
However, neither survives Bonferroni correction, with raw $p$ of $0.027$ on WHY and $0.52$ on WHAT, neither below the corrected threshold $\alpha{=}0.0167$.

The rank correlation cannot tighten further because models at similar pass@1 emit very different intent distributions.
At nearly identical pass@1 around $60\%$, WHY-fraction ranges from $18.1\%$ on Qwen3-235B-A22B to $40.7\%$ on GPT-5.4-Nano (per-model breakdown in Appendix~\ref{app:intent-per-model}).
A single rank correlation across the pool averages these opposite sub-populations into one number, and its small magnitude is consistent with that averaging-out.
We conclude that no robust intent signal was found, rather than that none exists: the 13-model pool bounds the detectable effect size, and bucket-level label reliability is mixed (WHY is reliable while OTHER is not; Appendix~\ref{app:kappa}).
This null motivates the controlled intervention of Section~\ref{sec:intervention} rather than establishing evidence of absence.

%% file: sections/04_intervention.tex
\section{Intervening on Comment Content}
\label{sec:intervention}

The observational evidence in Section~\ref{sec:observation} cannot determine whether self-emitted comments functionally affect pass@1.
We now run two controlled interventions.
In Section~\ref{sec:decomposition} we attach a comment block written by a stronger source model to the head of a recipient model's output.
We hold everything except the comment content fixed, isolating which property drives the change in pass@1.
The candidates are content correctness, line order, topic, and surface form.
Throughout, ``content'' refers to largely order-insensitive local claims about the solution rather than an ordered derivation, a scope the shuffled condition makes precise.
In Section~\ref{sec:self-elicit}, we ask whether the recipient itself can produce a comparable comment from its own prompt.
We test this with 14 models and 10 prompt variants, varying what we ask the recipient to write.

\subsection{What happens when we prefill the model with external comments?}
\label{sec:decomposition}

\input{tables/02_decomp_main}

\label{sec:decomp-setup}

We use four weak \textbf{prefix-recipients} (\texttt{Gemma-4-E4B-IT}, \texttt{Seed-Coder-8B-Instruct}, \texttt{Qwen3-8B}, \texttt{CodeGeeX4-9B}) and three stronger \textbf{prefix-sources}.
The sources are the two top performers in our self-run pool, \texttt{Grok-4.1-Fast} and \texttt{Gemini-3.1-Flash-Lite}, together with the state-of-the-art proprietary model \texttt{Claude-Opus-4.7}.
Each source writes \texttt{\#}-prefixed comments at the head of its solution to each LCB problem; we keep only its leading \texttt{\#}-block as the prefix attached to a recipient's output (extraction details and the source prompt in Appendix~\ref{app:prompt-prefix-source}).
We define all prefix conditions here.
\textbf{Source-written} attaches the source's comment block on all problems and is the union of \textbf{source-correct} (the subset where the source's own solution passes) and \textbf{source-wrong} (the subset where it fails); \textbf{shuffled} permutes the line order of source-correct blocks; \textbf{wrong-topic} attaches a comment block written for a different problem; \textbf{random-text} attaches length-matched random text (Appendix~\ref{app:decomp-random}).
Together they test content correctness, line order, topic, and surface form alone.
Decoding parameters are in Appendix~\ref{app:hyperparams}.

Every $\Delta$ in Table~\ref{tab:decomp-main} is measured against a \textbf{matched-prompt baseline}.
The baseline is the recipient's pass@1 on the cell's problem subset, run under the same prompt the source uses with no external comment attached.
This holds the prompt setting fixed and isolates the effect of \emph{which} comment block is attached.
Each cell is tested with a paired exact McNemar \citep{mcnemar1947} against this baseline; we apply Holm-Bonferroni correction \citep{holm1979} over $m{=}60$ tests at family-wise $\alpha{=}0.05$.
The 60 tests cover 12 cells each for source-written, source-correct, source-wrong, shuffled, and wrong-topic.

\label{sec:decomp-main}

\paragraph{Same-form comments lift the recipient only when their content is correct.}
Source-correct and source-wrong directly compare two blocks with the same form, differing only in whether the described solution is correct.
The two subsets are disjoint by construction, and each is tested against its own matched-prompt baseline to control for difficulty.
Correct content lifts the recipient's pass@1 by $17.2\%$ on average on the source-pass subset.
All 12 (recipient, source) source-correct cells pass Holm-Bonferroni at $p_\text{holm}{<}0.001$ (Table~\ref{tab:decomp-main}).
Same-form content describing a wrong solution leaves pass@1 essentially unchanged on the source-fail subset, and none of the 12 source-wrong cells reaches significance.
With form fixed and difficulty controlled at the subset level, the property that carries the lift is whether the comment describes a correct solution.
One residual confound is worth flagging.
Source-wrong comments are written on harder problems and may also differ from source-correct in tentative or confused style.
The matched-subset baselines control for problem difficulty in pass@1 terms, but equalizing the baseline level does not equalize headroom: recipients sit near the floor on the source-fail subset, so the source-wrong null cannot by itself separate wrong content from unsolvable problems.
We therefore treat source-wrong as supporting evidence, and rest the content claim on two comparisons that avoid the source-fail subset entirely: the corrupted-comment condition and the wrong-topic contrast below.

\paragraph{Corrupting one step of a correct comment removes the lift on the same problems.}
On the source-pass subset, we ask an LLM to make each correct comment block wrong in exactly one place, replacing the algorithm name or one key step while preserving length, line count, and tone (on average 2.3 of about 10 lines change; 50 samples verified by hand; prompt in Appendix~\ref{app:prompt-corruption}).
This holds difficulty, form, and style fixed while flipping only content correctness.
For all four recipients tested, the corrupted condition falls back to the baseline range, and the drop from source-correct is Holm-significant at $p_\text{holm}{<}0.001$ (Table~\ref{tab:corrupted}).

\input{tables/12_corrupted}

Style rewrites of the same correct blocks (paraphrased, compressed, hedged, vague) corroborate this: only removing specific task information hurts, and wording, length, and tone do not (Appendix~\ref{app:rewrites-table}).

\paragraph{Wrong-topic and random prefixes hurt rather than help.}
Two controls replace the useful content while preserving the comment-block form.
Wrong-topic comments use natural-language comments from a different problem, while length-matched random text removes coherent problem content entirely (Appendix~\ref{app:decomp-random}).
Both reduce recipient pass@1: wrong-topic by $20.8\%$ across 12 cells and random text by $17.9\%$ on average (Table~\ref{tab:decomp-random}).
Thus, the lift cannot be explained by merely placing extra tokens or comment-shaped text before the code.
Source-written and wrong-topic prefixes are closely matched in token count, comment-line count, and density, yet produce sharply different pass@1.
Coherent but topic-wrong comments hurt more than incoherent noise, suggesting that recipients tend to follow the content of the prefix rather than ignore it; a direct paired comparison confirms this asymmetry is significant (Appendix~\ref{app:decomp-random}).
The damage concentrates on easy problems (Table~\ref{tab:difficulty}), where the recipient would often have solved the task unaided.

\paragraph{Most of the source-correct lift survives line shuffling.}
We shuffle the source-written lines on each problem, preserving content and form but breaking order.
The recipient retains $68.0\%$ of the source-correct lift on average ($+11.7\%$ vs $+17.2\%$, Table~\ref{tab:decomp-main}), and 10 of 12 shuffled cells survive Holm correction at $p_\text{holm}{<}0.001$.
The remaining $32.0\%$ loss concentrates on easy problems where the recipient would have picked the right approach unaided (full per-recipient breakdown in Appendix~\ref{app:difficulty-detail}).
Our data is consistent with comment lines functioning as local hints, each contributing largely independently of the others with a smaller order-sensitive component.
They do not behave as a step-by-step derivation.
At the per-problem level, $89$ problems flip from fail to pass under correct prefill while only $3$ flip the other direction, a $30{:}1$ ratio (Appendix~\ref{app:flip-atlas}).
These 89 problems span all difficulty bands ($56$ medium, $18$ easy, $15$ hard), so the lift is not confined to rescuing the hardest problems. Qualitative examples are in Appendix~\ref{app:flip-examples}.

\paragraph{The content effect generalizes beyond Python single-file tasks to C\# and Java repository-level settings.}
\label{sec:generalization}

\input{tables/14_generalization}

LiveCodeBench is single-file Python competitive programming, so we replicate the core intervention on both splits of RepoClassBench \citep{deshpande2024repoclassbench}: C\# (61 tasks) and Java (130 tasks, excluding 2 whose build environment fails in every condition).
Each task asks the model to write a complete class from its natural-language description inside a real repository, and a generation passes only if it compiles against the repository and passes the repository's own tests.
Our strongest prefix-source, \texttt{Claude-Opus-4.7}, writes a comment block for each task; we inject it into all four prefix-recipients (Qwen3-8B, Seed-Coder-8B, CodeGeeX4-9B, Gemma-4-E4B-IT) under the baseline, source-written, and wrong-topic conditions of Section~\ref{sec:decomp-setup}, and test with paired exact McNemar as before.

In Table~\ref{tab:generalization}, every cell that reaches significance moves in the direction Section~\ref{sec:decomp-main} predicts, and no cell moves significantly against it; the remaining cells are flat within noise.
Correct comments lift pass@1 on Java when pooled across recipients (41 problems gained versus 16 lost, $p{=}0.001$), and wrong-topic comments hurt on both splits (Java: 18 gained versus 65 lost, $p{<}0.001$; C\#: 4 versus 16, $p{=}0.012$).
Qwen3-8B's Java wrong-topic cell is the only individually Holm-significant cell, moving $-13.8\%$ ($p_\text{holm}{=}0.022$); its Java source-correct cell shows the largest individual lift at $+16.7\%$ but does not survive per-cell correction alone, consistent with the effect emerging most clearly at the pooled level above.
On the same source-pass tasks, with task, model, and comment form all matched, the correct comment beats the wrong-topic comment 75 to 9 on Java and 20 to 2 on C\# (both $p{<}0.001$).

The effects concentrate where the recipient has headroom.
The C\# baselines sit below $12\%$, and Seed-Coder-8B starts from $66.7\%$ on the Java source-pass subset; in both regimes the per-cell deltas compress toward zero, echoing the difficulty analysis of Table~\ref{tab:difficulty}.
The content effect is therefore not an artifact of single-file Python competitive programming, although its visible size depends on how much room the recipient has to move.

\FloatBarrier
\subsection{Can prompting alone reproduce what prefilling external comments gives us?}
\label{sec:self-elicit}

\input{tables/03_grid_14x10}

We showed in Section~\ref{sec:decomposition} that a correct external comment block lifts the recipient's pass@1 substantially.
A natural follow-up is whether the recipient can produce a comparable comment from its own prompt, without an external source.
We test this with 14 models and 10 prompt variants.

The pool comprises the four prefix-recipients from Section~\ref{sec:decomposition} plus 10 popular self-run models spanning the pass tiers of Section~\ref{sec:form} (full list in Appendix~\ref{app:model-pool}).
Nine prompt policies vary along three dimensions; \textbf{base} (no policy) serves as the natural comparison anchor, for 10 conditions in total.

\emph{Volume} controls the amount of commenting the model produces:
\textbf{none} forbids all comments (inline, block, and docstrings),
\textbf{rare} caps comments at one short note per logical block and disallows docstrings or multi-line explanations,
\textbf{dense} requires a 2--5 line comment before each major block, with explicit reasoning about why the approach is correct.

\emph{Intent} controls what the comments say, mirroring the WHY/WHAT axis of Section~\ref{sec:intent}:
\textbf{WHAT} requires every block or non-trivial line to carry a comment describing what the code does,
\textbf{WHY} requires comments explaining the motivation behind each design choice,
\textbf{mix} requires both in the same comment.

\emph{Format} controls how the comments are organized:
\textbf{gp-doc} provides a human-convention reference --- the Google Python Style Guide \citep{gpstyleguide} that pairs per-function docstrings with sparse inline comments on non-obvious parts,
\textbf{top} reuses the prefix-source's prompt template from Section~\ref{sec:decomposition}, requesting a 5--15 line \texttt{\#}-prefixed comment block at the head of the recipient's output,
\textbf{inline} embeds reasoning as inline comments interleaved with the code, adapted from \citet{jiang2026thinkanywhere}.

The full policy text for each variant is in Appendix~\ref{app:prompts-variants}.
Figure~\ref{fig:variants-example} shows representative output under each variant on a common code body.

\begin{figure*}[t]
\centering
\makebox[\linewidth][c]{%
  \includegraphics[
    width=1.2\linewidth,
    trim=18 70 18 75,
    clip
  ]{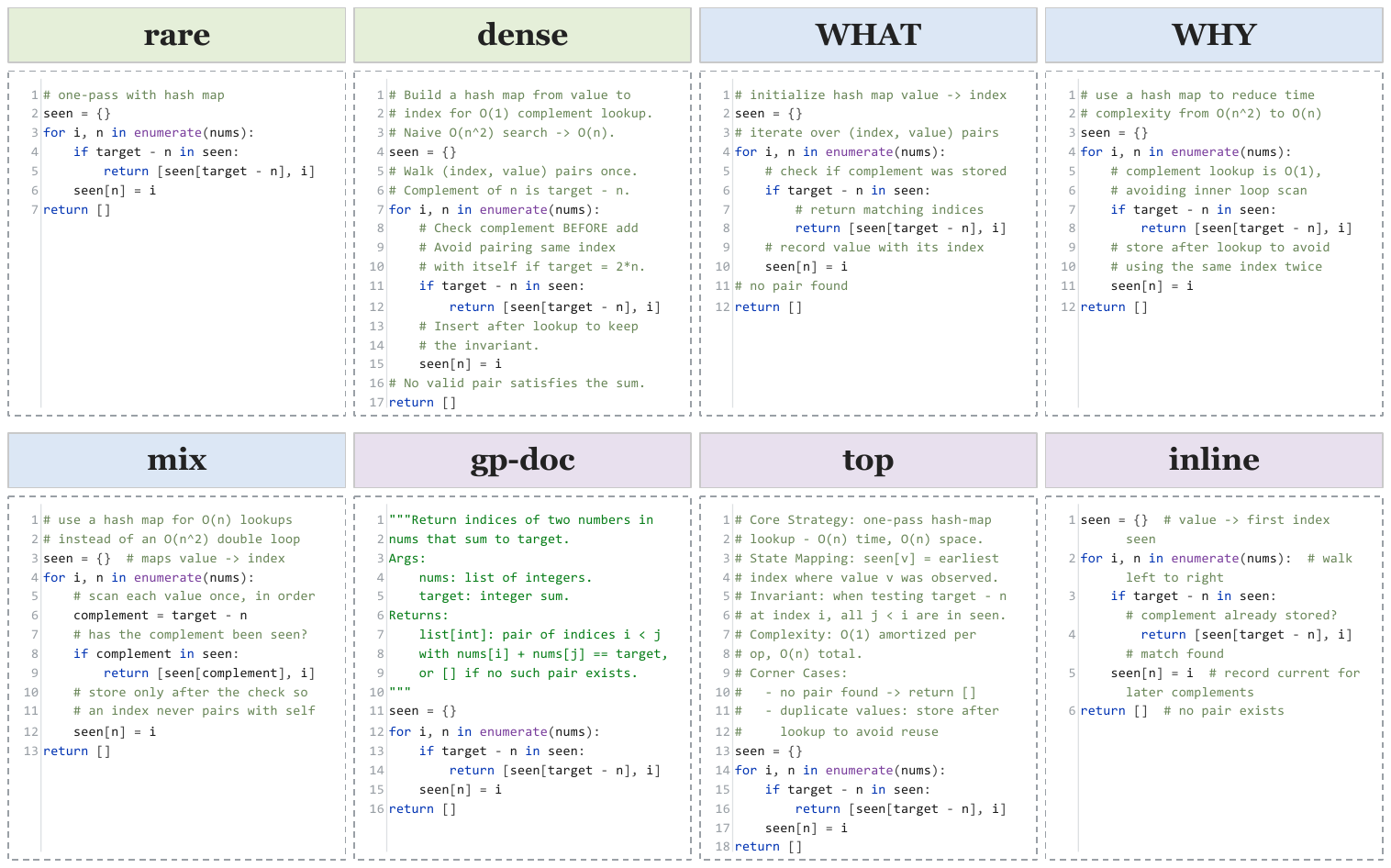}
}
\caption{Output under each prompt variant on the same Two-Sum body (a classic competitive-programming problem).}
\label{fig:variants-example}
\end{figure*}
Decoding follows Appendix~\ref{app:hyperparams}, with only the appended system policy varying across cells.
For each (model, variant) cell we report $\Delta$pass@1 against the model's \textbf{base} pass@1 on the same problems, paired-tested with McNemar and corrected with Holm-Bonferroni over the 126 non-baseline cells at family-wise $\alpha{=}0.05$.
We also report whether each model follows the surface instruction of its variant.

\label{sec:selfelicit-results}

\paragraph{Across the four recipients, self-elicitation produces at most a quarter of the external-comment lift.}
Among the four prefix-recipients from Section~\ref{sec:decomposition}, only Qwen3-8B under \textbf{dense} achieves a Holm-significant prompt-induced lift ($+4.4\%$).
For comparison, the same model gains $18.1\%$ on average when given an external correct comment block (Table~\ref{tab:decomp-main}).
Self-prompting therefore recovers only about $24\%$ of that effect.
For the other three recipients, no variant achieves a Holm-significant lift (Appendix~\ref{app:per-model-audit}).
A correct comment from a stronger model is therefore not interchangeable with a prompt asking the recipient to write its own.

\paragraph{The same prompt lifts some models and damages others.}
Every non-baseline variant in Table~\ref{tab:grid-14x10} has at least one model it lifts and at least one it damages.
The Holm-significant cells in Appendix~\ref{app:per-model-audit} include substantial moves in both directions for each variant.
GPT-5.1-Codex-Mini under \textbf{WHAT} gains $4.9\%$ ($p_\text{holm}=3.3 \times 10^{-3}$), while Gemma-4-E4B-IT under \textbf{none} loses $14.5\%$ ($p_\text{holm}=2.3 \times 10^{-30}$).
\textbf{top} lifts DeepSeek-V3.2 by $6.2\%$ but drops GPT-OSS-120B by $13.7\%$.
\textbf{none} lifts GPT-OSS-120B but is the largest damaging cell in the grid for Gemma-4-E4B-IT.
The variance is in the receiving model, not the prompt text.
A generic ``the prompt was unclear'' explanation does not fit, since the same instruction unlocks pass@1 on some models.

\paragraph{Compliance with the form instruction does not predict whether pass@1 goes up or down.}
Prompt instructions are followed in practice across all 10 variants, yet pass@1 still moves in both directions on each model.
The token shift under \textbf{none} does not explain the pass@1 changes.

\paragraph{Stronger self-planning prompts and a two-stage pipeline do not close the gap.}
Two further settings test whether the gap reflects weak prompting rather than a deeper limitation.
An explicit self-planning prompt (\textbf{SP}) asks the model to write a structured plan before coding, while a two-stage pipeline (\textbf{TS}) first elicits a standalone plan and then feeds it back to the same model before regeneration (full prompts in Appendix~\ref{app:prompt-sp-ts}).
We test both on two deliberately favorable models: \texttt{Gemini-3.1-Flash-Lite}, whose external comment blocks strongly help recipients, and \texttt{GPT-OSS-120B}, the grid model most responsive to intent-style prompts.
Across the four cells, pass@1 changes by only $+1.6\%$, $+1.9\%$, $-0.6\%$, and $+0.9\%$, with none significant.
Even the largest estimate recovers only about a tenth of the average external-comment effect.
Thus, the gap is not explained by weak prompting or the single-pass format: self-produced plans still do not reliably encode a correct solution.

\paragraph{Output-length changes do not explain the pass@1 effects.}
Under \textbf{none}, mean code tokens change by $-21.8\%$ across the 14-model pool, ranging from $-59\%$ on Gemma-4-E4B-IT to $+46\%$ on MiMo-V2-Flash (Appendix~\ref{app:per-cell-grid-form}).
Yet these large length shifts do not correspond to proportional pass@1 changes: Mistral-Small-2603 loses $43\%$ of its code tokens with no significant accuracy change, while MiMo-V2-Flash gains $46\%$ with none either.
The direction of the pass@1 effect instead depends on the receiving model.
This again supports the diagnosis from Section~\ref{sec:decomposition}: prompting controls comment form, but not whether the content encodes a correct solution.

\paragraph{Models that comment heavily at baseline are the most damaged by being told not to.}
The three biggest losses in the grid all come from \textbf{none} applied to models with a strong head-block commenting habit at baseline.
These are Gemma-4-E4B-IT ($-14.5\%$), GPT-5.4-Nano ($-13.6\%$), and Gemini-3.1-Flash-Lite ($-6.9\%$).
Each is the model's own largest single damage cell (Appendix~\ref{app:per-model-audit}).
The natural reading is that these models depend on the comment block their own decoding produces, and \textbf{none} removes it.
This is the cleanest cross-model pattern in the grid.
Intent-prompt effects do not partition the pool by baseline form alone.
GPT-OSS-120B and GPT-5.1-Codex-Mini gain from intent variants, but other low-comment models such as Llama-4-Maverick are damaged by the same variants.
The effect of each variant otherwise remains tied to the receiving model.

%% file: tables/02_decomp_main.tex
\begin{table*}[t]
  \centering
  \footnotesize
  \setlength{\tabcolsep}{4pt}
  \begin{tabular}{cccccccc}
  \toprule
  \textbf{Recipient} & \textbf{Source} & \textbf{Baseline} & \textbf{Source-written} & \textbf{Source-correct} & \textbf{Source-wrong} & \textbf{Shuffled} & \textbf{Wrong-topic} \\
  \midrule
  \multirow{3}{*}{Gemma-4-E4B-IT} & gemini & \multirow{3}{*}{54.2} & 64.3 (+9.9)$^*$ & 85.2 (+13.9)$^*$ & 10.5 ($-0.3$) & 76.2 (+4.9) & 46.6 ($-24.7$)$^*$ \\
  & claude &  & 62.5 (+6.6)$^*$  & 71.0 (+8.3)$^*$  & 3.9 ($-5.4$)  & 67.4 (+4.7) & 40.1 ($-22.5$)$^*$ \\
  & grok  &  & 69.9 (+14.1)$^*$ & 79.4 (+15.6)$^*$ & 10.1 (+4.3) & 74.0 (+10.2)$^*$ & 46.3 ($-17.5$)$^*$ \\
  \addlinespace
  \multirow{3}{*}{Qwen3-8B}  & gemini & \multirow{3}{*}{39.2} & 52.0 (+12.6)$^*$ & 70.8 (+17.2)$^*$ & 3.7 (+1.0) & 67.4 (+13.7)$^*$ & 19.6 ($-34.1$)$^*$ \\
  & claude &  & 53.3 (+12.8)$^*$ & 60.5 (+14.7)$^*$ & 3.1 (+0.0) & 58.4 (+12.6)$^*$ & 21.7 ($-24.1$)$^*$ \\
  & grok  &  & 60.5 (+19.7)$^*$ & 69.5 (+22.5)$^*$ & 3.6 (+2.2) & 63.6 (+16.6)$^*$ & 18.0 ($-29.0$)$^*$ \\
  \addlinespace
  \multirow{3}{*}{Seed-Coder-8B}  & gemini & \multirow{3}{*}{32.3} & 42.7 (+10.3)$^*$ & 58.3 (+14.4)$^*$ & 2.7 ($-0.3$) & 53.1 (+9.2)$^*$ & 26.3 ($-17.6$)$^*$ \\
  & claude &  & 44.5 (+11.1)$^*$ & 50.8 (+12.8)$^*$ & 0.8 ($-0.8$) & 48.3 (+10.3)$^*$ & 22.6 ($-15.4$)$^*$ \\
  & grok  &  & 49.4 (+15.7)$^*$ & 56.9 (+18.2)$^*$ & 2.2 (+0.0) & 48.9 (+10.1)$^*$ & 23.4 ($-15.3$)$^*$ \\
  \addlinespace
  \multirow{3}{*}{CodeGeeX4-9B}  & gemini & \multirow{3}{*}{19.4} & 38.1 (+18.6)$^*$ & 52.3 (+25.9)$^*$ & 1.7 (+0.0) & 45.4 (+19.0)$^*$ & 8.5 ($-18.0$)$^*$ \\
  & claude &  & 36.5 (+16.5)$^*$ & 41.5 (+18.7)$^*$ & 2.3 (+0.8) & 39.0 (+16.3)$^*$ & 6.5 ($-16.3$)$^*$ \\
  & grok  &  & 40.6 (+20.4)$^*$ & 46.8 (+23.8)$^*$ & 2.2 ($-0.7$) & 36.0 (+13.0)$^*$ & 7.7 ($-15.3$)$^*$ \\
  \bottomrule
  \addlinespace
  \multicolumn{8}{@{}p{0.97\textwidth}@{}}{\scriptsize $^*$ Holm-significant at $p_\text{holm}{<}0.001$ (paired McNemar, $m{=}60$). The 14 non-significant cells are all 12 source-wrong cells plus 2 Gemma shuffled cells (gemini, claude sources). Column definitions in Section~\ref{sec:decomp-setup}.} \\
  \end{tabular}
  \caption{Recipient pass@1 (\%) under each source-comment condition. Parentheses give $\Delta$ vs baseline.}
  \label{tab:decomp-main}
\end{table*}

%% file: tables/12_corrupted.tex
\begin{table}[t]
  \centering
  \footnotesize
  \setlength{\tabcolsep}{4pt}
  \resizebox{\linewidth}{!}{%
  \begin{tabular}{lccc}
  \toprule
  \textbf{Recipient} & \textbf{Baseline} & \textbf{Source-correct} & \textbf{Corrupted} \\
  \midrule
  Qwen3-8B & 53.6 & 70.8 (+17.2)$^*$ & 48.2 ($-5.4$)$^*$ \\
  Seed-Coder-8B & 43.9 & 58.3 (+14.4)$^*$ & 41.9 ($-2.0$) \\
  CodeGeeX4-9B & 26.4 & 52.3 (+25.9)$^*$ & 28.4 (+2.0) \\
  Gemma-4-E4B-IT & 71.3 & 85.2 (+13.9)$^*$ & 74.2 (+2.9) \\
  \bottomrule
  \end{tabular}}
  \vspace{0.3em}\par
  {\scriptsize\parbox{0.95\linewidth}{$^*$ Holm-significant vs baseline. The drop from source-correct to corrupted is Holm-significant at $p_\text{holm}{<}0.001$ for all four recipients.}}
  \caption{Recipient pass@1 (\%) on the source-pass subset (gemini source) when the correct comment block is minimally corrupted. Parentheses give $\Delta$ vs baseline.}
  \label{tab:corrupted}
\end{table}

%% file: tables/14_generalization.tex
\begin{table}[t]
  \centering
  \footnotesize
  \setlength{\tabcolsep}{4pt}
  \resizebox{\linewidth}{!}{%
  \begin{tabular}{llccc}
  \toprule
  \textbf{Recipient} & \textbf{Split} & \textbf{Source-correct} & \textbf{Source-wrong} & \textbf{Wrong-topic} \\
  \midrule
  \multirow{2}{*}{Qwen3-8B} & C\# & 19.2 (+0.0) & 5.7 (+2.9) & 0.0 ($-9.8$) \\
   & Java & 61.1 (+16.7) & 3.4 ($-3.4$) & 13.8 ($-13.8$)$^{*}$ \\
  \addlinespace
  \multirow{2}{*}{Seed-Coder-8B} & C\# & 38.5 (+15.4) & 2.9 (+0.0) & 9.8 ($-1.6$) \\
   & Java & 72.2 (+5.6) & 6.9 ($-5.2$) & 37.7 ($-4.6$) \\
  \addlinespace
  \multirow{2}{*}{CodeGeeX4-9B} & C\# & 15.4 (+0.0) & 0.0 ($-2.9$) & 4.9 ($-3.3$) \\
   & Java & 73.6 (+1.4) & 3.4 (+0.0) & 35.4 ($-6.2$) \\
  \addlinespace
  \multirow{2}{*}{Gemma-4-E4B-IT} & C\# & 34.6 (+15.4) & 2.9 (+0.0) & 4.9 ($-4.9$) \\
   & Java & 76.4 (+11.1) & 3.4 ($-6.9$) & 29.2 ($-11.5$) \\
  \bottomrule
  \end{tabular}}
  \vspace{0.3em}\par
  {\scriptsize\parbox{0.95\linewidth}{$^{*}$ Holm-significant at $\alpha{=}0.05$ over $m{=}24$ (recipient, split, condition) tests. Every cell that reaches significance moves in the direction of Section~\ref{sec:decomp-main}; no cell moves significantly against it.}}
  \caption{Recipient pass@1 (\%) on RepoClassBench (Claude-Opus-4.7 source). Parentheses give $\Delta$ vs the matched baseline; condition definitions follow Section~\ref{sec:decomp-setup}.}
  \label{tab:generalization}
\end{table}

%% file: tables/03_grid_14x10.tex
\begin{table*}[t]
  \centering
  \footnotesize
  \resizebox{\textwidth}{!}{%
  \begin{tabular}{ccccccccccc}
    \toprule
    & & \multicolumn{3}{c}{\textbf{Volume}} & \multicolumn{3}{c}{\textbf{Intent}} & \multicolumn{3}{c}{\textbf{Format}} \\
    \cmidrule(lr){3-5} \cmidrule(lr){6-8} \cmidrule(lr){9-11}
    \textbf{Model} & \textbf{base} & \textbf{none} & \textbf{rare} & \textbf{dense} & \textbf{WHAT} & \textbf{WHY} & \textbf{mix} & \textbf{gp-doc} & \textbf{top} & \textbf{inline} \\
    \midrule
    $\blacklozenge$ Grok-4.1-Fast            & 85.1 & 84.8 & 84.4 & 83.3 & 82.0 & 81.0$^*$ & 82.4 & 82.8 & 83.7 & 84.8 \\
    $\blacklozenge$ Gemini-3.1-Flash-Lite    & 72.5 & 65.6$^*$ & 67.0$^*$ & 72.4 & 69.8 & 71.6 & 70.7 & 72.3 & 66.7$^*$ & \textbf{72.7} \\
    GPT-OSS-120B                             & 69.1 & \textbf{72.5} & \textbf{71.5} & \textbf{70.0} & \textbf{74.3}$^*$ & \textbf{75.4}$^*$ & \textbf{76.3}$^*$ & \textbf{73.8} & 55.4$^*$ & 55.2$^*$ \\
    GPT-5.1-Codex-Mini                       & 64.6 & 64.2 & 64.4 & 56.5$^*$ & \textbf{69.5}$^*$ & \textbf{69.0}$^*$ & \textbf{68.9}$^*$ & \textbf{67.7} & 51.1$^*$ & 55.3$^*$ \\
    DeepSeek-V3.2                            & 64.3 & 58.2$^*$ & 58.8$^*$ & \textbf{65.9} & 62.7 & \textbf{66.7} & \textbf{65.7} & 59.2$^*$ & \textbf{70.4}$^*$ & 60.5 \\
    Qwen3-235B-A22B                          & 59.8 & 58.3 & 58.6 & \textbf{60.8} & \textbf{61.4} & \textbf{65.5}$^*$ & \textbf{63.2} & \textbf{61.6} & \textbf{61.3} & 58.7 \\
    GPT-5.4-Nano                             & 59.6 & 46.1$^*$ & 57.0 & \textbf{65.4}$^*$ & \textbf{63.0} & \textbf{63.3} & \textbf{62.3} & \textbf{61.4} & 57.6 & \textbf{64.8}$^*$ \\
    MiMo-V2-Flash                            & 55.9 & \textbf{56.3} & \textbf{56.2} & \textbf{61.6}$^*$ & \textbf{58.8} & \textbf{63.1}$^*$ & \textbf{58.9} & \textbf{57.9} & 46.9$^*$ & \textbf{60.2} \\
    Gemma-4-E4B-IT$^\dagger$                 & 53.7 & 39.2$^*$ & 46.1$^*$ & 42.2$^*$ & 47.4$^*$ & 52.5 & 50.9 & 49.7$^*$ & \textbf{54.2} & 52.2 \\
    Llama-4-Maverick                         & 45.3 & 43.5 & 45.2 & 44.6 & 32.6$^*$ & \textbf{45.4} & 42.9 & 38.5$^*$ & 33.7$^*$ & \textbf{45.6} \\
    Mistral-Small-2603                       & 44.2 & 42.5 & 42.9 & \textbf{48.5}$^*$ & 44.2 & \textbf{46.2} & \textbf{46.5} & \textbf{46.6} & \textbf{45.4} & \textbf{46.8} \\
    Seed-Coder-8B$^\dagger$                  & 33.4 & \textbf{35.4} & \textbf{33.9} & 29.5 & \textbf{35.1} & 31.7 & \textbf{34.2} & 31.5 & 32.3 & \textbf{36.1} \\
    Qwen3-8B$^\dagger$                       & 37.0 & 35.4 & 36.4 & \textbf{41.3}$^*$ & \textbf{39.7} & \textbf{40.1} & \textbf{38.5} & \textbf{39.2} & \textbf{39.2} & \textbf{39.6} \\
    CodeGeeX4-9B$^\dagger$                   & 22.1 & 20.2 & 20.3 & 19.8 & 17.7$^*$ & 17.1$^*$ & 18.6 & 20.7 & 19.4 & 21.5 \\
    \bottomrule
  \end{tabular}}
  \vspace{0.3em}\par
  {\scriptsize\parbox{0.97\textwidth}{$^*$ Holm-significant at $\alpha{=}0.05$ (paired McNemar, $m{=}126$; 38/126 survive). $\blacklozenge$ = prefix-source; $^\dagger$ = prefix-recipient. Variant prompts in Appendix~\ref{app:prompts-variants}.}}
  \caption{Effect of 10 prompt variants on pass@1 across 14 models. \textbf{Bold} = exceeds \textbf{base}.}
  \label{tab:grid-14x10}
\end{table*}

%% file: sections/05_conclusion.tex
\section{Conclusion}
\label{sec:conclusion}

In this paper, we asked whether comments, natural-language text an LLM emits in the same forward pass as the code it generates, functionally support its own pass@1.
The natural distribution of commenting habits across 88 models on LiveCodeBench does not track pass@1, whether we measure how much models comment or what their comments say.
A controlled intervention then attached a comment block from a stronger source to a weaker recipient and varied what the source's comment said while holding everything else fixed.
Correct content raises the recipient's pass@1 by $17.2\%$ on average across 12 such pairings, while comments matched in surface form but describing a wrong solution leave it unchanged, and coherent comments on a different problem damage it by $20.8\%$.
The lift therefore comes from the correctness of the described solution.

Asking the recipient to produce the same comments from its own prompt fails for three of the four recipients, and recovers at most about $24\%$ of the external-comment lift in the best case.
Prompt engineering controls how the comments look but not whether their content is correct.
A comment helps pass@1 only when its content describes a correct solution, which a stronger source can provide but a recipient cannot reliably elicit through its own prompt.

Practically, a model's reliance on comments can be checked cheaply with the none-versus-base contrast, and comment policies should follow that profile: suppressing comments costs up to $14.5\%$ on comment-dependent models, while helping others.
Closing the gap itself likely requires training-time methods, which we leave to future work.

%% file: sections/limitations.tex
\section*{Limitations}
\label{sec:limitations}
Although our intervention isolates content correctness as the property that drives the comment-induced lift, the experimental scope remains limited.
Despite testing ten prompt variants plus explicit self-planning and two-stage pipelines (Section~\ref{sec:self-elicit}), we did not find a prompting method that reliably reproduces the external-comment lift through self-elicitation; closing this gap likely requires training-time methods, which we leave to future work.

%% file: sections/ethics.tex
\section*{Ethics Statement}
\label{sec:ethics}

This work is an empirical study of code generated by publicly available LLMs on a public benchmark (LiveCodeBench-v6) under its public release terms. 
No human subjects research was conducted. 
The 100-snippet calibration set used to validate the LLM judge was hand-labelled by one of the authors. 
Self-run generations are produced via publicly available APIs and open-weights models with their corresponding licences. 
Total compute is approximately \$500 USD of API spend plus 200 A100-hours of local GPU time. The authors used commercial LLM assistants for sentence-level writing revision and analysis-script drafting, while all experimental design, statistical claims, and final manuscript content remain the authors' own. 

%% file: sections/acknowledgments.tex
\section*{Acknowledgments}

This research / project is supported by the Australian Research Council under its Discovery Early Career Researcher Award (DECRA) funding scheme (DE260100192), and the National Research Foundation, under its Investigatorship Grant (NRF-NRFI08-2022-0002). Any opinions, findings and conclusions or recommendations expressed in this material are those of the author(s) and do not reflect the views of National Research Foundation, Singapore.

%% file: appendix/A_stats.tex
\section{Statistical details}
\label{app:stats}

\subsection{LLM-judge calibration on 100-snippet stratified subset}
\label{app:kappa}

\paragraph{Seven-class codebook.}
The human reference annotator and the LLM judge in Section~\ref{sec:intent} both apply a seven-class codebook to each in-code comment line.
\textbf{rationale} explains WHY a design choice was made (algorithm trade-off, edge case justification, motivation).
\textbf{algorithm} names or summarises the high-level approach (``binary search'', ``two-pointer'', ``DP with memoisation'').
\textbf{step} describes what a specific line or local block of code does in procedural terms.
\textbf{mechanism} explains a specific code mechanism, invariant, or low-level detail.
\textbf{definition} labels or names a variable, data structure, or quantity.
\textbf{decoration} covers metadata, TODOs, section headers, and aesthetic comments.
\textbf{misleading} is factually incorrect with respect to the code it accompanies.
The main-text WHY / WHAT / OTHER aggregation maps \texttt{rationale} to WHY, \texttt{algorithm + step + mechanism + definition} to WHAT, and \texttt{decoration + misleading} to OTHER.

\paragraph{Calibration.}
We hand-labelled 100 in-code comment snippets stratified by intent class and compared the human annotation against the LLM judge (\texttt{Gemini-3.1-Flash-Lite}) on the same snippets.
For each of the seven intent classes, we compute the per-label binary $\kappa$ (does the human / does the judge assign this label).

\begin{table}[h]
  \centering
  \small
  \begin{tabular}{lrr}
    \toprule
    Label & Binary $\kappa$ & 95\% bootstrap CI \\
    \midrule
    \texttt{rationale}  & $+0.74$ & $[+0.58, +0.87]$ \\
    \texttt{algorithm}  & $+0.72$ & $[+0.55, +0.86]$ \\
    \texttt{mechanism}  & $+0.38$ & $[+0.18, +0.57]$ \\
    \texttt{definition} & $+0.48$ & $[+0.27, +0.66]$ \\
    \texttt{step}       & $+0.02$ & $[-0.12, +0.16]$ \\
    \texttt{decoration} & $+0.11$ & $[-0.04, +0.27]$ \\
    \texttt{misleading} & $\sim 0$ & ($n{=}2$, too sparse) \\
    \bottomrule
  \end{tabular}
  \caption{Per-label binary Cohen's $\kappa$ on the 100-snippet calibration subset. The two anchor labels (\texttt{rationale} for WHY, \texttt{algorithm} for WHAT) show $\kappa \geq 0.7$; the remaining five labels have lower agreement and are absorbed into WHY/WHAT/OTHER.}
  \label{tab:kappa-per-label}
\end{table}

\paragraph{Bucket-level agreement.}
Because Section~\ref{sec:intent} draws its conclusions on the aggregated WHY/WHAT/OTHER buckets, we also report agreement at the bucket level on the same 100 snippets (Table~\ref{tab:kappa-bucket}).
WHAT's positive rate is extreme (85--88\% of snippets), a regime where Cohen's $\kappa$ understates agreement \citep{feinstein1990kappa}, so we additionally report the prevalence-robust Gwet's AC1 \citep{gwet2008ac1}.
WHY, the bucket that carries the only directional trend in Section~\ref{sec:intent}, is reliable ($\kappa = \text{AC1} = 0.74$).
WHAT's low $\kappa$ is consistent with the high-prevalence $\kappa$ paradox, and raw agreement and AC1 remain high.
OTHER does not reach usable reliability, so we exclude it from quantitative statements.

\begin{table}[h]
  \centering
  \footnotesize
  \setlength{\tabcolsep}{3pt}
  \begin{tabular}{lcccc}
    \toprule
    Bucket & Pos.\ rate (H/J) & Raw agr. & Cohen's $\kappa$ & AC1 \\
    \midrule
    WHY   & 0.46 / 0.43 & 0.87 & $+0.74$ & 0.74 \\
    WHAT  & 0.85 / 0.88 & 0.85 & $+0.36$ & 0.80 \\
    OTHER & 0.36 / 0.22 & 0.62 & $+0.10$ & 0.35 \\
    \bottomrule
  \end{tabular}
  \caption{Bucket-level human-vs-judge agreement on the 100-snippet calibration subset. 95\% bootstrap CIs: WHY $[+0.59,+0.86]$, WHAT $[+0.07,+0.61]$, OTHER $[-0.09,+0.29]$.}
  \label{tab:kappa-bucket}
\end{table}

\paragraph{Multiple-comparison correction for the intent--pass@1 correlations.}
In Section~\ref{sec:intent} we test three Spearman correlations in parallel (WHY, WHAT, OTHER versus pass@1).
Because these are tested simultaneously we apply Bonferroni correction with $\alpha = 0.05 / 3 = 0.0167$ per test, which bounds the family-wise false-positive rate at $0.05$.
None of the three raw $p$-values falls below this corrected threshold.

\subsection{Per-model intent fractions}
\label{app:intent-per-model}

Table~\ref{tab:intent-per-model} reports the per-model 7-class intent fractions for each of the 13 models in the Section~\ref{sec:intent} pool, with the WHY / WHAT / OTHER groupings shown as supercolumns.

\input{tables/04_intent_per_model}

%% file: tables/04_intent_per_model.tex
\begin{table*}[h]
  \centering
  \small
  \begin{tabular}{lcccccccc}
    \toprule
    & & \textbf{WHY} & \multicolumn{4}{c}{\textbf{WHAT}} & \multicolumn{2}{c}{\textbf{OTHER}} \\
    \cmidrule(lr){3-3} \cmidrule(lr){4-7} \cmidrule(lr){8-9}
    \textbf{Model} & \textbf{pass@1} & \texttt{rat.} & \texttt{alg.} & \texttt{step} & \texttt{mech.} & \texttt{def.} & \texttt{dec.} & \texttt{mis.} \\
    \midrule
    Gemini-3.1-FL   & 72.5 & 35.3 &  9.8 &  7.6 & 30.1 & 14.5 &  2.7 & 0.0 \\
    GPT-OSS-120B    & 69.1 & 12.0 &  5.0 &  7.0 & 40.2 & 29.1 &  6.7 & 0.0 \\
    DeepSeek-V3.2   & 64.3 & 29.8 & 14.0 &  9.5 & 29.8 & 12.4 &  4.6 & 0.0 \\
    Qwen3-235B      & 59.8 & 18.1 & 11.7 & 13.5 & 44.2 &  9.3 &  3.3 & 0.0 \\
    GPT-5.4-Nano    & 59.6 & 40.7 & 14.6 &  3.9 & 23.1 & 14.2 &  3.5 & 0.0 \\
    gpt-4.1-mini    & 58.9 & 29.3 & 12.8 &  6.9 & 35.4 & 13.4 &  2.1 & 0.0 \\
    MiMo-V2-Flash   & 55.9 & 19.2 &  8.8 & 10.0 & 43.5 & 13.9 &  4.4 & 0.1 \\
    gpt-oss-20b     & 55.7 & 11.9 &  4.9 &  7.3 & 39.1 & 25.4 & 11.4 & 0.0 \\
    Gemini-2.5-FL   & 51.3 & 23.9 &  5.8 &  9.9 & 32.9 & 10.1 & 17.2 & 0.2 \\
    Llama-4-Mav     & 45.3 &  7.4 &  3.9 & 10.4 & 38.4 &  9.9 & 29.9 & 0.1 \\
    Mistral-S-2603  & 44.2 & 17.6 & 12.8 & 12.5 & 29.3 &  6.6 & 20.3 & 0.7 \\
    Qwen3-Coder-30B & 38.5 & 12.5 &  7.5 & 20.3 & 46.9 &  8.7 &  4.1 & 0.1 \\
    DSCoder-V2-Lite & 25.2 &  3.5 &  7.9 & 39.2 & 36.3 &  7.4 &  5.5 & 0.2 \\
    \bottomrule
  \end{tabular}
  \caption{Per-model 7-class intent fractions (\%) across the 13-model Section~\ref{sec:intent} pool. Column abbreviations: \texttt{rat.} = rationale, \texttt{alg.} = algorithm, \texttt{mech.} = mechanism, \texttt{def.} = definition, \texttt{dec.} = decoration, \texttt{mis.} = misleading. The three groupings used in Section~\ref{sec:intent} are: \textbf{WHY} = \texttt{rat.}; \textbf{WHAT} = \texttt{alg.} + \texttt{step} + \texttt{mech.} + \texttt{def.}; \textbf{OTHER} = \texttt{dec.} + \texttt{mis.}. Fractions are averaged over each model's comment-bearing samples on LiveCodeBench, sorted by pass@1 descending. Rows sum to 100\%.}
  \label{tab:intent-per-model}
\end{table*}

%% file: appendix/B_prompts.tex
\section{Prompts}
\label{app:prompts}

All inference uses the decoding parameters in Appendix~\ref{app:hyperparams}; only the prompt content differs across experiments.

\subsection{Prefix-source generation prompt}
\label{app:prompt-prefix-source}

Used in Section~\ref{sec:decomposition} for all three prefix-sources (\texttt{Grok-4.1-Fast}, \texttt{Gemini-3.1-Flash-Lite}, \texttt{Claude-Opus-4.7}).

\paragraph{System.}
\begin{quote}\small\ttfamily
You are an expert competitive programmer.
\end{quote}

\paragraph{User template (with \texttt{\{question\}} = LCB problem statement; \texttt{\{starter\_code\_block\}} = optional starter code or empty string).}
\begin{quote}\small\ttfamily
Solve the following programming problem. Output ONE Python code block containing:

1. First, 5--15 lines of `\#`-prefixed comments at the top of the function body (or top of the script if no function), explaining your algorithm: high-level approach, key observations, edge cases, complexity.

2. Then, the actual implementation that solves the problem.

Format requirements:
- Output exactly one ```python ... ``` block, nothing else.
- Prefixed comment lines MUST be `\#`-prefixed Python comments at the top.
- Prefixed comment lines MUST appear before any executable code.
- The implementation must be self-contained and correctly solve the problem.

PROBLEM:
\{question\}

\{starter\_code\_block\}

OUTPUT:
\end{quote}

Decoding parameters follow Appendix~\ref{app:hyperparams}, with the added API-side flag \texttt{extra\_body=\{"include\_reasoning": True\}} so that reasoning-enabled models (Opus 4.7, Gemini 3.1) still emit visible content.

\paragraph{Block extraction.}
From each source generation we keep only the contiguous block of \texttt{\#}-prefixed lines at the top of the first Python code block.
Any executable code or imports below those lines is discarded, so the prefix-recipient sees a pure natural-language prefix matched in surface form to a comment block.

\subsection{LLM-judge prompt (intent classification, Section~\ref{sec:intent})}
\label{app:prompt-judge}

The judge is \texttt{Gemini-3.1-Flash-Lite} (deterministic, called once per code snippet).

\paragraph{System.}
\begin{quote}\small\ttfamily
You are an annotation assistant for an empirical study of AI-generated code comments.

Your task is to classify the intent of each comment line in model-generated code. Each comment line maps to EXACTLY ONE intent (single-label, mutually exclusive).

Rules:
1. Focus only on comment intent.
2. Use the provided intent labels only.
3. For each intent label, output a non-negative integer count of comment lines that fall into that label.
4. Each comment line counts in EXACTLY ONE intent. If a comment fits multiple labels, apply the tie-break priority below and pick the highest.
5. Therefore: total comment lines in the snippet == rationale + algorithm + step + mechanism + definition + decoration + misleading.
6. If there are no meaningful comments at all, set "none" to 1 and all other intent counts to 0. If "none" is positive, every other intent count must be 0.
7. Output exactly one JSON object and nothing else.

Tie-break priority (when a comment fits multiple intents, pick the highest):
  misleading > rationale > algorithm > step > mechanism > definition > decoration
\end{quote}

\paragraph{User template (\texttt{\{CODE\}} = code snippet to annotate).}
\begin{quote}\small\ttfamily
Analyze the intent of each comment line in the following code output. Each line is assigned to EXACTLY ONE intent.

Return exactly one JSON object with these fields:
- rationale
- algorithm
- step
- mechanism
- definition
- decoration
- misleading
- none
- evidence

[\ldots full label definitions and tie-break priority repeated here\ldots]

Code:
\{CODE\}
\end{quote}

Full label definitions follow the seven-class codebook in Section~\ref{sec:intent}.

\subsection{Prefix-recipient base prompt (Section~\ref{sec:decomposition} and Section~\ref{sec:self-elicit})}
\label{app:prompt-prefix-recipient}

Prefix-recipients use the standard LCB system prompt and user template.

\paragraph{System (generic).}
\begin{quote}\small\ttfamily
You are an expert Python programmer. You will be given a question (problem specification) and will generate a correct Python program that matches the specification and passes all tests.
\end{quote}

\paragraph{User (for stdin/stdout problems).}
\begin{quote}\small\ttfamily
\#\#\# Question:

\{question\_content\}

\#\#\# Format: Read the inputs from stdin solve the problem and write the answer to stdout (do not directly test on the sample inputs). Enclose your code within delimiters as follows. Ensure that when the python program runs, it reads the inputs, runs the algorithm and writes output to STDOUT.

```python
\# YOUR CODE HERE
```

\#\#\# Answer: (use the provided format with backticks)
\end{quote}

\paragraph{Prefix-injection mechanism (Section~\ref{sec:decomposition}).}
The injection conditions in Section~\ref{sec:decomposition} use \texttt{tokenizer.apply\_chat\_template(messages, continue\_final\_message=True)}: the user prompt is unchanged, and a new \texttt{assistant}-role message is appended containing the prefix-source comments plus a truncated \texttt{```python\ldots} fence; the prefix-recipient continues from this in-place truncation.
This is implemented in \texttt{render\_chat\_prompt\_with\_prefix\_support}; the post-generation stitch back into pass-evaluatable code is implemented in \texttt{merge\_lcb\_assistant\_continuations}.

\paragraph{Variant policy mechanism (Section~\ref{sec:self-elicit}).}
The 10 self-elicitation variants in Section~\ref{sec:self-elicit} do not use prefix injection.
Each variant is a policy string appended to the system prompt above; the \textbf{base} variant is the empty string (LCB default).

\subsection{Self-elicitation variant policies (full text)}
\label{app:prompts-variants}

Below are the full policy strings for the 10 variants reported in Table~\ref{tab:grid-14x10}.
The main text uses descriptive short tags throughout (\textbf{base}, \textbf{none}, \textbf{rare}, \textbf{dense}, \textbf{WHAT}, \textbf{WHY}, \textbf{mix}, \textbf{gp-doc}, \textbf{top}, \textbf{inline}).
Each policy is appended verbatim to the LCB system prompt (Appendix~\ref{app:prompt-prefix-recipient}); the \textbf{base} variant is the empty string (LCB default).

\paragraph{\textbf{base} (LCB baseline).}
Empty string -- the system prompt is unchanged.

\paragraph{\textbf{none} (comment-free).}
\begin{quote}\small\ttfamily
COMMENT POLICY:
- Output only the final code, with all comments removed.
- Do not include inline comments, block comments, docstrings, or any explanatory text.
\end{quote}

\paragraph{\textbf{rare} (sparse comments).}
\begin{quote}\small\ttfamily
COMMENT POLICY:
- Comments are allowed but must be SHORT and rare.
- At most 1 short comment per logical block.
- Do NOT add docstrings or multi-line explanation comments.
- Do NOT repeat the same idea across multiple comments.
\end{quote}

\paragraph{\textbf{dense} (detailed multi-line comments).}
\begin{quote}\small\ttfamily
COMMENT POLICY:
- Add detailed comments explaining reasoning and edge cases.
- Before each major block, include a multi-line comment (2--5 lines).
- Comments should explain why the approach is correct.
- Still output ONLY the code (no markdown fences, no extra text).
\end{quote}

\paragraph{\textbf{WHAT} (WHAT-intent).}
\begin{quote}\small\ttfamily
COMMENT POLICY (WHAT -- intent):
You are an expert programmer. When generating code, you must add inline comments that explain WHAT each part of the code does.

Requirements for comments:
- Every logical block or non-trivial line must have a comment
- Comments should describe the functionality and purpose of the code
- Use the format: \# [what this code does]
- Comments must be concise and precise

Example:
\# initialize a dictionary to store character frequencies
freq = \{\}
\# iterate through each character in the string
for c in s:
    freq[c] = freq.get(c, 0) + 1
\end{quote}

\paragraph{\textbf{WHY} (WHY-intent).}
\begin{quote}\small\ttfamily
COMMENT POLICY (WHY -- motivation):
You are an expert programmer. When generating code, you must add inline comments that explain WHY you made each design decision.

Requirements for comments:
- Focus on the reasoning and motivation behind implementation choices
- Explain why this approach was chosen over alternatives
- Highlight key constraints or trade-offs considered
- Use the format: \# [why this decision was made]
\end{quote}

\paragraph{\textbf{mix} (WHAT + WHY mixed).}
\begin{quote}\small\ttfamily
COMMENT POLICY (WHAT + WHY -- mixed):
You are an expert programmer. When generating code, you must add inline comments that explain both WHAT the code does and WHY you made each design decision.

Requirements:
- Describe the functionality of each logical block (what)
- Explain the reasoning behind key implementation choices (why)
- For critical decisions, explicitly mention alternatives that were considered and rejected
- Use the format: \# [what] + [why if non-trivial]
\end{quote}

\paragraph{\textbf{gp-doc} (Google Python Style Guide).}
\begin{quote}\small\ttfamily
COMMENT AND DOCSTRING POLICY (Google Python Style Guide):
You are an expert Python programmer who strictly follows the Google Python Style Guide for comments and docstrings (Section 3.8).

Block and inline comments:
- Only comment tricky or non-obvious parts.
- Put a short comment block before complicated operations to explain intent.
- Use inline comments only for non-obvious behavior.
- Focus comments on WHAT you are trying to achieve and the intent, not how.

Docstrings:
- Every function must have a docstring using triple double quotes.
- First line is a one-line summary (\textless= 80 chars) ending with a period.
- After a blank line, include Args / Returns / Raises sections when relevant.
\end{quote}

\paragraph{\textbf{top} (reasoning-at-top, mirrors the prefix-source prompt).}
This variant uses the same prompt template as the prefix-source generation in Appendix~\ref{app:prompt-prefix-source}, so that a model's self-elicited \texttt{top} output is the directly-comparable analogue of the prefix that a stronger model would have produced for the same problem.

\paragraph{\textbf{inline} (inline reasoning, adapted from \citealp{jiang2026thinkanywhere}).}
The wording of this variant is adapted from \citet{jiang2026thinkanywhere}, who study on-demand inline reasoning at arbitrary token positions during code generation.
\begin{quote}\small\ttfamily
You need to generate code with inline reasoning embedded as comments.

Generate code where your reasoning process is expressed as inline comments within the code itself.

Rules:
1. Reasoning MUST appear as inline comments (\# or //) embedded within the code
2. Comments MUST appear immediately before or alongside the code they reason about
3. The code must remain valid and executable
4. Do not output any text outside the code block
\end{quote}

\subsection{Self-planning (SP) and two-stage (TS) prompts (Section~\ref{sec:self-elicit})}
\label{app:prompt-sp-ts}

Both conditions append to or replace the same LCB system prompt used throughout (Appendix~\ref{app:prompt-prefix-recipient}); \textbf{SP} is single-pass, \textbf{TS} is a two-call pipeline that elicits a standalone plan before code.

\paragraph{\textbf{SP} (self-planning policy, appended to the system prompt).}
\begin{quote}\small\ttfamily
Before writing any code, produce an explicit solution plan as \# comment lines at the top of your code block: restate the core difficulty, choose an algorithm and justify why it is correct for the constraints, decompose the implementation into numbered steps, and list edge cases you will handle. Only after completing this written plan, implement the code that follows it step by step.
\end{quote}

\paragraph{\textbf{TS} stage 1 (plan-only user prompt; no code).}
\begin{quote}\small\ttfamily
Write a detailed SOLUTION PLAN for the following problem. Do NOT write any code. The plan must: restate the core difficulty, choose an algorithm and justify its correctness against the constraints, decompose the solution into numbered implementation steps, and list the edge cases to handle.

\#\#\# Question:
\{question\}

\#\#\# Plan:
\end{quote}

\paragraph{\textbf{TS} stage 2.}
The stage-1 plan is inserted verbatim into the standard LCB user prompt immediately before the answer marker (\texttt{"\#\#\# Your plan (written by you in a previous step):\textbackslash n\{plan\}\textbackslash n\textbackslash n\#\#\# Answer:"}), and the model implements from there with no further system-prompt change.

\subsection{Corrupted-comment prompt (Section~\ref{sec:decomposition})}
\label{app:prompt-corruption}

Used to produce the corrupted condition of Table~\ref{tab:corrupted}. The corruptor is \texttt{Gemini-3.1-Flash-Lite} at temperature 0 with JSON output; inputs are the correct comment blocks on the source-pass subset.

\paragraph{System (abridged; requirement lists shown in full).}
\begin{quote}\small\ttfamily
You are assisting a controlled experiment on code comments. Your task is MINIMAL SEMANTIC CORRUPTION: given a programming problem and a comment block that correctly describes a working solution, edit the comment block so that it describes an INCORRECT solution to this problem, while changing as little of the surface text as possible.

1. CORRUPT THE CONTENT. Change exactly ONE central algorithmic commitment so the described approach would NOT solve the problem correctly: replace the core algorithm/technique with one that fails for this problem, OR invert/break one key step, condition, or invariant the solution relies on. The corrupted approach must be genuinely wrong for THIS problem, not an alternative valid solution.

2. MINIMAL EDIT. Rewrite only the lines that state or directly depend on the corrupted commitment. Copy every other line VERBATIM. Aim to keep at least half of the lines unchanged.

3. PRESERVE ALL SURFACE PROPERTIES: same number of lines and indentation; each edited line within roughly $\pm$20\% of the original length; same confident, declarative tone (never ``maybe'', ``I think''); same specificity (keep concrete names, quantities, and complexity claims, updated to match the corrupted approach); same phrasing patterns.

4. STAY PLAUSIBLE. The corrupted block must read like a competent programmer confidently describing a reasonable-sounding approach.
\end{quote}

The model returns the corrupted block together with one-sentence statements of the removed and introduced commitments and the number of lines edited; we verify a difficulty-stratified sample of 50 corruptions by hand.

\subsection{Style-rewrite prompts (Section~\ref{sec:decomposition})}
\label{app:prompt-rewrites}

Used to produce the four rewrites of Table~\ref{tab:rewrites}, sharing one system skeleton whose invariants require the rewritten block to describe the same solution (same algorithm, steps, conditions, and invariants) and to change exactly one surface property. The per-condition modulation instructions are:

\begin{description}\small
\item[paraphrased] Reword every line with different vocabulary and sentence structure but the same meaning; same line count, tone, and specificity.
\item[compressed] Reduce the block to roughly 40--60\% of its character count while remaining information-lossless at the level of algorithmic commitments.
\item[hedged] Keep every claim identical but rewrite the tone from confident to tentative, weaving hedges (``maybe'', ``I think'', ``probably'') across most lines.
\item[vague] Replace concrete commitments with generic descriptions at every line (algorithm names $\rightarrow$ broad category, exact conditions $\rightarrow$ qualitative phrasing, complexity claims $\rightarrow$ vague assurances), without introducing any incorrect claim: the block becomes less informative, not wrong.
\end{description}

Rewrites failing an automatic validation of these surface constraints are discarded, giving the per-condition $n$ in Table~\ref{tab:rewrites}.

\subsection{RepoClassBench replication setup (Section~\ref{sec:generalization})}
\label{app:generalization-setup}

Tasks come from the C\# and Java splits of RepoClassBench \citep{deshpande2024repoclassbench}; each provides a natural-language class description, a target file inside a real repository, and the repository's own build and test harness (dotnet/xunit for C\#, Maven/JUnit for Java).
The recipient prompt contains the class description, the file path, and the original file's import/namespace header; the generated class is written back into the repository, and a generation passes only if the repository compiles and the task's test class passes.
The prefix-source (\texttt{Claude-Opus-4.7}) is prompted as in Appendix~\ref{app:prompt-prefix-source} with \texttt{//}-prefixed comments; recipients receive the extracted comment block through the same assistant-prefill mechanism as Section~\ref{sec:decomposition}.
Two Java tasks whose Maven build fails in every condition, including baseline, are excluded.
Pooled contrasts use paired exact McNemar over the union of (task, recipient) pairs.

%% file: appendix/C_models_compute.tex
\section{Models and compute}
\label{app:models-compute}

\subsection{Model pool}
\label{app:model-pool}

\paragraph{Self-run by us (Section~\ref{sec:form}; 14 models).}
Ten models run via the OpenRouter API with homogeneous decoding (listed by base pass@1 descending):
\begin{itemize}\itemsep0pt
  \item \texttt{Grok-4.1-Fast}
  \item \texttt{Gemini-3.1-Flash-Lite}
  \item \texttt{GPT-OSS-120B}
  \item \texttt{GPT-5.1-Codex-Mini}
  \item \texttt{DeepSeek-V3.2}
  \item \texttt{Qwen3-235B-A22B}
  \item \texttt{GPT-5.4-Nano}
  \item \texttt{MiMo-V2-Flash}
  \item \texttt{Llama-4-Maverick}
  \item \texttt{Mistral-Small-2603}
\end{itemize}
Four additional weak models run locally with vLLM (\texttt{trust\_remote\_code=True}, \texttt{dtype=bfloat16}, \texttt{max\_model\_len=16384}, \texttt{gpu\_memory\_utilization=0.8}, \texttt{enforce\_eager=True}); these same four also serve as the prefix-recipients in Section~\ref{sec:decomposition} (listed by base pass@1 descending):
\begin{itemize}\itemsep0pt
  \item \texttt{google/gemma-4-E4B-it}
  \item \texttt{Qwen/Qwen3-8B}
  \item \texttt{ByteDance-Seed/Seed-Coder-8B-Instruct}
  \item \texttt{zai-org/codegeex4-all-9b}
\end{itemize}

\paragraph{Leaderboard submissions (Section~\ref{sec:form}; 74 models).}
LCB leaderboard submission generations as published by \citet{lcb_submissions}; the full per-model identifiers and form metrics appear in Tables~\ref{tab:form-profile}--\ref{tab:form-profile-2}.

\paragraph{Intent-analysis pool (Section~\ref{sec:intent}; 13 models).}
8 of the 10 self-run OpenRouter models above (those that produce enough comments per problem for reliable per-model intent estimation), plus 5 LCB leaderboard submissions selected to broaden pass-spectrum coverage of the LLM-judge call:
\begin{itemize}\itemsep0pt
  \item \texttt{GPT-4.1-Mini-2025-04-14}
  \item \texttt{Qwen3-Coder-30B-A3B-Instruct}
  \item \texttt{Gemini-2.5-Flash-Lite}
  \item \texttt{GPT-OSS-20B}
  \item \texttt{DeepSeek-Coder-V2-Lite-Instruct}
\end{itemize}

\paragraph{Prefix-sources used in Section~\ref{sec:decomposition} (3 strong models, OpenRouter API).}
\begin{itemize}\itemsep0pt
  \item \texttt{x-ai/grok-4.1-fast}
  \item \texttt{google/gemini-3.1-flash-lite-preview}
  \item \texttt{anthropic/claude-opus-4.7}
\end{itemize}

\paragraph{Self-elicit pool used in Section~\ref{sec:self-elicit} (14 models, listed by base pass@1 descending).}
\begin{itemize}\itemsep0pt
  \item \texttt{Grok-4.1-Fast}, \texttt{Gemini-3.1-Flash-Lite} (also prefix-sources)
  \item \texttt{GPT-OSS-120B}, \texttt{GPT-5.1-Codex-Mini}, \texttt{DeepSeek-V3.2}, \texttt{Qwen3-235B-A22B}, \texttt{GPT-5.4-Nano}, \texttt{MiMo-V2-Flash}, \texttt{Llama-4-Maverick}, \texttt{Mistral-Small-2603}
  \item \texttt{Gemma-4-E4B-IT}, \texttt{Qwen3-8B}, \texttt{Seed-Coder-8B-Instruct}, \texttt{CodeGeeX4-9B} (also prefix-recipients)
\end{itemize}

\subsection{Decoding hyperparameters}
\label{app:hyperparams}

Table~\ref{tab:hyperparams} summarises the decoding parameters across all experiments.

\begin{table}[h]
  \centering
  \small
  \begin{tabular}{p{2.5cm}lp{2.7cm}}
    \toprule
    Parameter & Value & Where \\
    \midrule
    $T$ (temperature)         & 0       & all \\
    $n$ (samples / problem)   & 1       & all \\
    \texttt{max\_tokens}      & 4096    & §\ref{sec:decomposition}, §\ref{sec:self-elicit} \\
    \texttt{top\_p}           & 0.95    & prefix-source API \\
    \texttt{max\_model\_len}  & 16384   & vLLM recipients \\
    \texttt{dtype}            & bfloat16 & vLLM recipients \\
    \texttt{trust\_remote\_code} & True & vLLM recipients \\
    \texttt{enforce\_eager}   & True    & vLLM recipients \\
    Judge MIME                & JSON & §\ref{sec:intent} judge \\
    Judge \texttt{include\_reasoning} & True & §\ref{sec:decomposition} API \\
    \bottomrule
  \end{tabular}
  \caption{Decoding hyperparameters across experiments. All experiments use
  $n{=}1$ and $T{=}0$; the only variable is the prompt content.}
  \label{tab:hyperparams}
\end{table}

\subsection{Compute statement}
\label{app:compute}

\paragraph{Local GPU.}
All prefix-recipient inference (Section~\ref{sec:decomposition}) and self-elicit grid (Section~\ref{sec:self-elicit}) ran on a single NVIDIA A100 80GB with 240\,GB host memory and 13 CPU workers.
Total: approximately 200 A100-hours across all generations.

\paragraph{API spend.}
Prefix-source generation (\texttt{Grok-4.1-Fast}, \texttt{Gemini-3.1-Flash-Lite}, \texttt{Claude-Opus-4.7}) on the full LCB-v6 used approximately \$500 USD of OpenRouter credit.

\paragraph{Reproducibility.}
Re-analysis (recomputing metrics from the existing generations) requires no GPU or API spend; re-running the controlled-decomposition section end-to-end requires the A100 budget above plus access to the listed OpenRouter models.

%% file: appendix/D_supplementary.tex
\section{Supplementary results}
\label{app:supplementary}


\subsection{Style-rewrite results}
\label{app:rewrites-table}
\input{tables/13_rewrites}

\subsection{Form metric definitions}
\label{app:form-metrics}

Each of the 88 model outputs is parsed under one shared tokenizer (\texttt{cl100k\_base}) so that comment volumes are comparable across vendors, since native tokenizers are not available for all leaderboard submissions.
The four form metrics are defined as follows.
\textbf{tok-cmt\%} is the fraction of output tokens that fall inside a comment, where ``inside a comment'' covers the span from a comment marker (\texttt{\#}, or a triple-quoted string used as a docstring) to the end of that comment.
\textbf{block\%}, \textbf{inline\%}, and \textbf{doc\%} follow the PEP-8 partition of comment forms \citep{pep8}.
\textbf{block\%} is the share of comment tokens that are full-line \texttt{\#} comments.
\textbf{inline\%} is the share of comment tokens that are \texttt{\#} comments appearing on the same line as code.
\textbf{doc\%} is the share of comment tokens that are docstrings.
These three shares sum to 100\% over the subset of samples that contain at least one comment.

\subsection{Full 88-model form profile}
\label{app:full-form-profile}

Table~\ref{tab:toptier-spread} reports the form spread of the 16 top-tier models at matched pass@1.
\input{tables/01_toptier_spread}
Tables~\ref{tab:form-profile}--\ref{tab:form-profile-2} report the full per-model form profile for all 88 models, sorted by pass@1 descending.

\input{tables/05_form_profile}

\subsection{Within-model $\rho$ vs pass@1}
\label{app:within-model-rho}

Table~\ref{tab:within-model-rho} reports the within-model Spearman~$\rho$ between each form axis and per-problem pass@1, computed independently for each of the 88 models in the pool.
On three of the four axes (\texttt{block\%}, \texttt{inline\%}, \texttt{doc\%}) no model exceeds $|\rho|{=}0.32$ and at most $13.0\%$ exceed $|\rho|{=}0.20$.
Comment volume (\texttt{tok-cmt\%}) has a non-trivial upper tail: $22.0\%$ of models exceed $|\rho|{=}0.20$ and the maximum reaches $|\rho|{=}0.53$, which still corresponds to under $30.0\%$ explained variance.
A minority of models therefore do exhibit within-model variation, but never at a magnitude that would change the Section~\ref{sec:form} conclusion.

\input{tables/06_within_model_rho}


\subsection{Length-matched random-text control}
\label{app:decomp-random}

Table~\ref{tab:decomp-random} reports the recipient's pass@1 under the length-matched random-text control (\texttt{\#}-prefixed filler matched in length to the source-written block).
The control is run on all three sources' source-pass subsets (gemini/claude/grok, $n{=}757/892/870$) across four recipients (12 cells), mirroring the wrong-topic design in Table~\ref{tab:decomp-main}; the matched-prompt baseline is the recipient's pass@1 on the same subset (the same per-cell baseline convention as Table~\ref{tab:decomp-main}).
Random text damages pass@1 on all four recipients (mean $\Delta{=}{-}17.9\%$ across 12 cells), but the damage is smaller than the wrong-topic damage reported in Table~\ref{tab:decomp-main} (mean $\Delta{=}{-}20.8\%$ across 12 cells).
We test this asymmetry directly with a paired comparison between the wrong-topic and random-text pass vectors on common problems, per (recipient, source) cell with Holm correction over 12 cells, plus a pooled test.
Pooled over 10{,}076 paired observations, 946 problems pass only under random text versus 650 only under wrong-topic ($p{<}0.001$).
Per cell, 5 of 12 are individually Holm-significant, all in the wrong-topic-worse direction, and no cell reverses, though the per-cell effect is heterogeneous.

\input{tables/07_decomp_random}

\subsection{Difficulty stratification: per-prefix-recipient table}
\label{app:difficulty-detail}

Table~\ref{tab:difficulty} reports the per-(recipient, source, condition) breakdown across LCB difficulty bands.
Each row's $\Delta$ is computed against the matched-prompt baseline on the corresponding (cell, difficulty) sub-subset.
Source-correct prefills lift pass@1 most on medium problems for most (recipient, source) pairs; wrong-topic and random-text damage concentrates on easy problems.

\input{tables/08_difficulty}

\subsection{Per-problem flip atlas}
\label{app:flip-atlas}

The flip atlas is computed across the 4 prefix-recipients under the source-correct (grok prefix-source) condition vs the matched-prompt baseline, on the 870-problem source-pass subset.
Of these, 89 ($10.2\%$) are \textbf{source-transferable} (at least 3 of the 4 prefix-recipients flip baseline-fail to source-correct-pass, with 0 loss in the opposing direction), 3 ($0.3\%$) are \textbf{source-toxic} (at least 3 flip the other way with 0 gain), and 778 ($89.4\%$) are mixed or neutral.

\paragraph{Difficulty distribution of source-transferable problems.}
Among the 89 source-transferable problems: 56 medium, 18 easy, 15 hard.

\subsection{Per-problem flip atlas: qualitative anchors}
\label{app:flip-examples}

Among the 119 problems where 6 or more of 12 voters lift unanimously (and no voter loses), the prefix-source's comment block typically reroutes the prefix-recipient from a plausible-but-wrong strategy to the correct one.

\paragraph{Example: LeetCode \texttt{3617} (easy, 12/12 voters lift).}
Prefix-source: Gemini-3.1-Flash-Lite. Prefix-recipient: Qwen3-8B.

The prefix-source's comment block (injected as prefix-recipient prefix) reads:
\begin{quote}\small\ttfamily
\# The problem asks for the number of possible original strings given that
\# at most one character group could have been typed with extra repetitions.
\# 1. Identify all contiguous groups of identical characters.
\# 2. For each group of length L $>$ 1, there are (L-1) ways to have
\#    extra characters in this specific group.
\# 3. Total possibilities = 1 (the word itself) + sum of (L\_i - 1) for all groups.
\# 4. This simplifies to: 1 + sum(L\_i - 1) for all groups.
\end{quote}

Without the prefix, Qwen3-8B's baseline attempt \emph{multiplies} the group counts (a plausible-looking but incorrect approach derived from the structure of the problem):
\begin{quote}\small\ttfamily
def possibleStringCount(self, word):\\
\hspace*{1em}total = 1\\
\hspace*{1em}\ldots count consecutive groups\ldots\\
\hspace*{1em}total *= count   \# wrong: product, not sum\\
\hspace*{1em}return total
\end{quote}

With the prefix, Qwen3-8B implements the prefix-source's stated formula (sum of group-lengths minus 1, plus 1) and passes:
\begin{quote}\small\ttfamily
total = 1 + sum(g - 1 for g in groups)
\end{quote}

The prefix-source's prefix supplies a single strategic decision -- ``it's a sum, not a product'' -- that the prefix-recipient was unable to find on its own.
Across the 119 strong-transferable problems, this pattern repeats: the prefix-source's correct algorithmic choice is reflected verbatim into the prefix-recipient's solution, while the prefix-recipient absent prefix would commit to one of several morphologically similar but incorrect strategies.


\subsection{McNemar $+$ Holm-Bonferroni grid (full statistics for Section~\ref{sec:self-elicit})}
\label{app:mcnemar-holm}

Of the 126 (model, variant) non-baseline cells in Table~\ref{tab:grid-14x10}, 38 survive Holm--Bonferroni at $\alpha{=}0.05$; these are highlighted in Tables~\ref{tab:grid-14x10} and~\ref{tab:per-model-audit}.
The strongest signals are:
\begin{itemize}\itemsep0pt
  \item \texttt{Gemma-4-E4B-IT} \textbf{none}: $\Delta{=}{-}14.5\%$, $p_\text{holm}=2.3 \times 10^{-30}$
  \item \texttt{GPT-OSS-120B} \textbf{inline}: $\Delta{=}{-}13.9\%$, $p_\text{holm}=2.1 \times 10^{-25}$
  \item \texttt{GPT-OSS-120B} \textbf{top}: $\Delta{=}{-}13.7\%$, $p_\text{holm}=9.3 \times 10^{-29}$
  \item \texttt{GPT-5.4-Nano} \textbf{none}: $\Delta{=}{-}13.6\%$, $p_\text{holm}=4.3 \times 10^{-20}$
  \item \texttt{GPT-OSS-120B} \textbf{mix}: $\Delta{=}{+}7.2\%$, $p_\text{holm}=2.3 \times 10^{-6}$
\end{itemize}

Paired-bootstrap 95\% confidence intervals \citep{efron1979bootstrap} (1000 resamples on problem ID) have median width $4.4\%$.
All cells with $|\Delta| \geq 6\%$ survive Holm correction.

\subsection{Self-elicit grid: per-model dramatic-cell audit}
\label{app:per-model-audit}

Table~\ref{tab:per-model-audit} reports the full 14-row per-model audit with baseline form profile, the largest- and smallest-$\Delta$ variant per model (bold where Holm-significant), and significant-cell count per model.

\input{tables/09_per_model_audit}

\subsection{Self-elicit grid: variant compliance details}
\label{app:variant-compliance}

Table~\ref{tab:variant-compliance} reports per-variant mean output code tokens and mean tok-cmt\% aggregated across the 14 models (column means of Table~\ref{tab:per-cell-grid-form}).
Models reliably follow the form instruction at the population level (\textbf{none} drops mean tok-cmt\% from $22.7\%$ to $6.3\%$; \textbf{dense}, \textbf{WHY}, \textbf{mix}, \textbf{top}, \textbf{inline} all push it above $50\%$), yet the pass@1 effect is decoupled from compliance (Section~\ref{sec:selfelicit-results}).

\input{tables/10_variant_compliance}

\subsection{Per-cell mean code tokens and tok-cmt\%}
\label{app:per-cell-grid-form}

Table~\ref{tab:per-cell-grid-form} reports the per-cell mean output code tokens and tok-cmt\% across the 14$\times$10 self-elicit grid; this is the per-(model, variant) detail behind the per-variant aggregates in Table~\ref{tab:variant-compliance}.

Two patterns stand out.
First, form policies are followed in aggregate: \textbf{none} pushes tok-cmt\% below $10\%$ on 12 of 14 models (Qwen3-235B at $26.6\%$ and MiMo-V2-Flash at $24.6\%$ remain partially compliant), and \textbf{dense}, \textbf{WHY}, and \textbf{mix} push tok-cmt\% above $40\%$ on 13 of 14 models (one model exception per variant: Seed-Coder-8B under \textbf{dense} at $37.2\%$, Llama-4-Maverick under \textbf{WHY} at $34.4\%$ and under \textbf{mix} at $26.0\%$).
Second, the token shift under \textbf{none} is highly model-dependent: mean output code tokens shift by an average of $-21.8\%$ across the 14 models, with the range spanning Gemma-4-E4B-IT contracting from 1067 to 437 tokens ($-59\%$) up to MiMo-V2-Flash \emph{expanding} from 1119 to 1628 tokens ($+46\%$) --- the only model whose mean tokens grow under \textbf{none}, because the lost comment scaffolding lets the model fail to stop on its longest tail of generations.
The verbose form variants (\textbf{top}, \textbf{dense}, \textbf{inline}) inflate mean tokens to $0.7$--$2.6{\times}$ the \textbf{base} value, peaking at Mistral-Small-2603 under \textbf{top} (365 to 953, $2.6{\times}$) and DeepSeek-V3.2 under \textbf{top} (603 to 1094, $1.8{\times}$); the lower end comes from MiMo-V2-Flash under \textbf{inline} (1119 to 767, $0.7{\times}$).
These shifts describe how strictly each model complies with the \emph{form} of the prompt; they do not predict the corresponding pass@1 effect (Section~\ref{sec:selfelicit-results}).

\input{tables/11_per_cell_grid_form}

\subsection{Per-model output token shift under \textbf{none}}
\label{app:none-token-drop}

Table~\ref{tab:none-token-drop} reports the per-model mean output code token count under \textbf{base} and \textbf{none}, measured under a shared \texttt{cl100k\_base} tokenizer over all 1055 LCB problems.
The token shift is asymmetric: 13 of 14 models contract under \textbf{none} (largest drop: Gemma-4-E4B-IT, $-59\%$), but MiMo-V2-Flash actually \emph{expands} by $+46\%$.

\begin{table*}[h]
  \centering
  \setlength{\tabcolsep}{4pt}
  \begin{tabular}{lrrrc}
    \toprule
    \textbf{Model} & \textbf{base} & \textbf{none} & \textbf{shift \%} & \textbf{pass sig?} \\
    \midrule
    Grok-4.1-Fast           & 277  & 267  & $-3.6$  & --         \\
    Gemini-3.1-Flash-Lite   & 441  & 235  & $-46.5$ & \checkmark \\
    GPT-OSS-120B            & 225  & 183  & $-18.4$ & \checkmark \\
    GPT-5.1-Codex-Mini      & 154  & 130  & $-15.9$ & --         \\
    DeepSeek-V3.2           & 603  & 269  & $-55.4$ & \checkmark \\
    Qwen3-235B-A22B         & 495  & 429  & $-13.3$ & --         \\
    GPT-5.4-Nano            & 679  & 486  & $-28.4$ & \checkmark \\
    MiMo-V2-Flash           & 1119 & 1628 & $+45.5$ & --         \\
    Gemma-4-E4B-IT          & 1067 & 437  & $-59.0$ & \checkmark \\
    Llama-4-Maverick        & 239  & 193  & $-19.5$ & --         \\
    Mistral-Small-2603      & 365  & 207  & $-43.2$ & --         \\
    Qwen3-8B                & 399  & 347  & $-13.2$ & --         \\
    Seed-Coder-8B           & 207  & 170  & $-17.6$ & --         \\
    CodeGeeX4-9B            & 210  & 177  & $-15.9$ & --         \\
    \bottomrule
  \end{tabular}
  \caption{Per-model mean output code tokens under \textbf{base} vs \textbf{none}, sorted by base pass@1 descending. Tokens count code only (markdown fences, surrounding explanation, and \texttt{<think>} tags excluded). The pass-sig column marks whether \textbf{none}'s pass@1 effect survives Holm-Bonferroni in Appendix~\ref{app:per-model-audit}.}
  \label{tab:none-token-drop}
\end{table*}

%% file: tables/13_rewrites.tex
\begin{table}[t]
  \centering
  \footnotesize
  \setlength{\tabcolsep}{5pt}
  \begin{tabular}{lcc}
  \toprule
  \textbf{Rewrite (axis)} & \textbf{Qwen3-8B} & \textbf{Seed-Coder-8B} \\
  \midrule
  paraphrased (wording) & $-1.0$ & $-1.0$ \\
  compressed (length) & $-2.7$ & $+3.4^{*}$ \\
  hedged (confidence) & $-0.3$ & $+1.4$ \\
  vague (specificity) & $-7.0^{***}$ & $-5.3^{**}$ \\
  \bottomrule
  \addlinespace
  \multicolumn{3}{@{}p{0.47\textwidth}@{}}{\scriptsize $^{*}p{<}0.05$, $^{**}p{<}0.01$, $^{***}p{<}0.001$; unmarked cells are not significant.} \\
  \end{tabular}
  \caption{Change in pass@1 (\%) vs source-correct under four single-axis style rewrites of the same correct comment blocks (gemini source; $n{=}691/700/724/618$ rewrites pass validation per condition).}
  \label{tab:rewrites}
\end{table}

%% file: tables/01_toptier_spread.tex
\begin{table*}[t]
  \centering
  \footnotesize
  \setlength{\tabcolsep}{6pt}
  \begin{tabular}{lccccc}
    \toprule
    \textbf{Model} & \textbf{pass@1} & \textbf{tok-cmt\%} & \textbf{block\%} & \textbf{inline\%} & \textbf{doc\%} \\
    \midrule
    DeepSeek-R1-0528             & 84.4 & 0.1  & 33.8 & 66.2 & 0.0  \\
    EXAONE-4.0-32B               & 80.9 & 0.1  & 55.0 & 45.0 & 0.0  \\
    OpenReasoning-Nemotron-32B   & 81.0 & 0.1  & 58.6 & 41.4 & 0.0  \\
    Grok-4.1-Fast                & 85.1 & 1.4  & 90.4 & 9.6  & 0.0  \\
    Kimi-k1.6-IOI                & 80.2 & 7.7  & 81.3 & 18.7 & 0.0  \\
    XBai-o4-medium               & 80.1 & 8.5  & 89.1 & 10.9 & 0.0  \\
    Kimi-k1.6-IOI-high           & 86.0 & 10.7 & 87.2 & 12.8 & 0.0  \\
    QwQ-Max-Preview              & 80.0 & 12.5 & 89.8 & 10.2 & 0.0  \\
    Qwen3-235B-A22B              & 80.4 & 13.5 & 90.8 & 9.2  & 0.0  \\
    O4-Mini (High)               & 87.3 & 28.2 & 95.2 & 3.6  & 1.2  \\
    O4-Mini (Medium)             & 84.5 & 28.9 & 94.8 & 4.0  & 1.2  \\
    O3 (High)                    & 84.7 & 32.4 & 48.1 & 25.0 & 26.8 \\
    O1-2024-12-17 (High)         & 83.2 & 50.1 & 81.7 & 3.6  & 14.6 \\
    Gemini-2.5-Pro-06-05         & 84.3 & 50.4 & 80.7 & 1.4  & 17.9 \\
    Gemini-2.5-Pro-05-06         & 82.8 & 53.6 & 89.4 & 9.3  & 1.3  \\
    Gemini-2.5-Pro-03-25         & 81.5 & 71.4 & 78.0 & 4.3  & 17.7 \\
    \bottomrule
  \end{tabular}
  \caption{Comment form across the 16 LCB models with pass@1 $\geq 80.0\%$, sorted by \texttt{tok-cmt\%}.}
  \label{tab:toptier-spread}
\end{table*}

%% file: tables/05_form_profile.tex
\begin{table*}[t]
  \centering
  \footnotesize
  \setlength{\tabcolsep}{4pt}
  \begin{tabular}{lcccccc}
    \toprule
    \textbf{Model} & \textbf{Source} & \textbf{pass@1} & \textbf{tok-cmt\%} & \textbf{block\%} & \textbf{inline\%} & \textbf{doc\%} \\
    \midrule
    O4-Mini (High) & sub & 87.3 & 28.2 & 95.2 & 3.6 & 1.2 \\
    Kimi-k1.6-IOI-high & sub & 86.0 & 10.7 & 87.2 & 12.8 & 0.0 \\
    Grok-4.1-Fast & sr & 85.1 & 1.4 & 90.4 & 9.6 & 0.0 \\
    O3 (High) & sub & 84.7 & 32.4 & 48.1 & 25.0 & 26.8 \\
    O4-Mini (Medium) & sub & 84.5 & 28.9 & 94.8 & 4.0 & 1.2 \\
    DeepSeek-R1-0528 & sub & 84.4 & 0.1 & 33.8 & 66.2 & 0.0 \\
    Gemini-2.5-Pro-06-05 & sub & 84.3 & 50.4 & 80.7 & 1.4 & 17.9 \\
    O1-2024-12-17 (High) & sub & 83.2 & 50.1 & 81.7 & 3.6 & 14.6 \\
    Gemini-2.5-Pro-05-06 & sub & 82.8 & 53.6 & 89.4 & 9.3 & 1.3 \\
    Gemini-2.5-Pro-03-25 & sub & 81.5 & 71.4 & 78.0 & 4.3 & 17.7 \\
    OpenReasoning-Nemotron-32B & sub & 81.0 & 0.1 & 58.6 & 41.4 & 0.0 \\
    EXAONE-4.0-32B & sub & 80.9 & 0.1 & 55.0 & 45.0 & 0.0 \\
    Qwen3-235B-A22B & sub & 80.4 & 13.5 & 90.8 & 9.2 & 0.0 \\
    Kimi-k1.6-IOI & sub & 80.2 & 7.7 & 81.3 & 18.7 & 0.0 \\
    XBai-o4-medium & sub & 80.1 & 8.5 & 89.1 & 10.9 & 0.0 \\
    QwQ-Max-Preview & sub & 80.0 & 12.5 & 89.8 & 10.2 & 0.0 \\
    Grok-3-Mini (High) & sub & 78.4 & 18.4 & 75.5 & 24.5 & 0.0 \\
    O1-2024-12-17 (Med) & sub & 78.3 & 56.8 & 76.7 & 3.0 & 20.3 \\
    DeepSeek-R1-Preview & sub & 77.9 & 8.4 & 85.8 & 13.7 & 0.5 \\
    Llama-3-1-Nemotron-Ultra-253B-v1 & sub & 77.7 & 4.6 & 92.0 & 8.0 & 0.0 \\
    O3-Mini-2025-01-31 (High) & sub & 77.7 & 45.4 & 94.4 & 5.5 & 0.2 \\
    O4-Mini (Low) & sub & 77.4 & 29.9 & 93.8 & 4.0 & 2.1 \\
    Gemini-2.5-Flash-05-20 & sub & 76.2 & 60.4 & 91.9 & 5.4 & 2.7 \\
    O1-2024-12-17 (Low) & sub & 75.9 & 58.5 & 82.4 & 2.6 & 14.9 \\
    O3-Mini-2025-01-31 (Med) & sub & 75.4 & 46.1 & 94.3 & 5.5 & 0.1 \\
    Gemini-2.5-Flash-04-17 & sub & 75.1 & 66.2 & 92.4 & 4.1 & 3.5 \\
    DeepCoder-14B-Preview & sub & 73.3 & 23.5 & 94.0 & 6.0 & 0.0 \\
    Gemini-3.1-Flash-Lite & sr & 72.5 & 39.5 & 97.1 & 0.3 & 2.7 \\
    O3-Mini-2025-01-31 (Low) & sub & 70.6 & 45.3 & 94.8 & 5.0 & 0.2 \\
    Claude-Opus-4 (Thinking) & sub & 70.4 & 20.6 & 90.4 & 8.7 & 0.8 \\
    GPT-OSS-120B & sr & 69.1 & 15.7 & 71.4 & 23.5 & 5.0 \\
    Claude-Sonnet-4 (Thinking) & sub & 68.5 & 15.9 & 85.8 & 13.3 & 0.9 \\
    O1-Mini-2024-09-12 & sub & 68.4 & 9.1 & 93.0 & 6.6 & 0.3 \\
    GPT-5.1-Codex-Mini & sr & 64.6 & 0.6 & 88.5 & 11.5 & 0.0 \\
    Llama-3-1-Nemotron-Nano-8B-v1 & sub & 64.4 & 3.9 & 96.4 & 3.6 & 0.0 \\
    DeepSeek-V3.2 & sr & 64.3 & 22.6 & 99.1 & 0.9 & 0.0 \\
    Claude-3.7-Sonnet & sub & 63.5 & 23.9 & 81.4 & 14.9 & 3.7 \\
    DeepSeek-R1-Lite-Preview & sub & 63.1 & 13.4 & 91.3 & 8.7 & 0.0 \\
    Claude-Opus-4 & sub & 62.4 & 26.8 & 93.0 & 5.9 & 1.1 \\
    QwQ-32B-Preview & sub & 59.9 & 12.5 & 89.4 & 10.4 & 0.2 \\
    Qwen3-235B-A22B & sr & 59.8 & 37.7 & 96.8 & 3.0 & 0.3 \\
    GPT-5.4-Nano & sr & 59.6 & 39.4 & 97.8 & 2.1 & 0.1 \\
    \bottomrule
  \end{tabular}
  \caption{Full per-model form profile on LCB (part 1 of 2: top 42 of 88 models, sorted by pass@1). \textbf{sr} = self-run (10 OpenRouter + 4 vLLM); \textbf{sub} = LCB leaderboard submission. Form metrics defined in Appendix~\ref{app:form-metrics}.}
  \label{tab:form-profile}
\end{table*}

\begin{table*}[t]
  \centering
  \footnotesize
  \setlength{\tabcolsep}{4pt}
  \begin{tabular}{lcccccc}
    \toprule
    \textbf{Model} & \textbf{Source} & \textbf{pass@1} & \textbf{tok-cmt\%} & \textbf{block\%} & \textbf{inline\%} & \textbf{doc\%} \\
    \midrule
    Claude-Sonnet-4 & sub & 59.4 & 24.6 & 88.7 & 10.6 & 0.7 \\
    gpt-4.1-mini-2025-04-14 & sub & 58.9 & 50.2 & 98.4 & 1.6 & 0.0 \\
    Gemini-Flash-2.0-Thinking-12-19 & sub & 56.5 & 30.2 & 90.2 & 5.5 & 4.3 \\
    MiMo-V2-Flash & sr & 55.9 & 36.3 & 99.5 & 0.4 & 0.1 \\
    gpt-oss-20b & sub & 55.7 & 20.8 & 59.4 & 13.0 & 27.6 \\
    Gemini-Flash-2.0-Thinking-01-21 & sub & 55.7 & 30.3 & 90.1 & 6.3 & 3.6 \\
    O1-Preview-2024-09-12 & sub & 55.6 & 12.4 & 78.8 & 21.2 & 0.0 \\
    MetaStone-L1-7B & sub & 54.1 & 7.0 & 87.3 & 12.7 & 0.0 \\
    Gemma-4-E4B-IT & sr & 53.7 & 58.6 & 96.5 & 0.8 & 2.8 \\
    gemini-2.5-flash-lite & sub & 51.3 & 38.4 & 96.2 & 3.6 & 0.2 \\
    Gemini-Flash-2.0-Thinking & sub & 51.2 & 3.0 & 76.4 & 23.1 & 0.5 \\
    DeepSeek-V3 & sub & 51.0 & 31.0 & 98.5 & 1.5 & 0.0 \\
    Gemini-Exp-1206 & sub & 50.0 & 0.1 & 71.9 & 28.1 & 0.0 \\
    Qwen2.5-Ins-72B & sub & 49.1 & 8.8 & 94.9 & 5.1 & 0.0 \\
    Claude-3.5-Sonnet-20241022 & sub & 48.7 & 20.3 & 93.4 & 6.5 & 0.1 \\
    Qwen2.5-Coder-Ins-32B & sub & 48.3 & 9.4 & 96.1 & 3.9 & 0.0 \\
    Claude-3.5-Sonnet-20240620 & sub & 48.0 & 7.7 & 89.7 & 10.3 & 0.0 \\
    Qwen2.5-Ins-32B & sub & 47.1 & 6.7 & 93.7 & 6.3 & 0.0 \\
    Llama-4-Maverick & sr & 45.3 & 11.3 & 86.5 & 5.0 & 8.5 \\
    Mistral-Small-2603 & sr & 44.2 & 16.9 & 98.7 & 1.3 & 0.0 \\
    GPT-4O-2024-05-13 & sub & 43.4 & 12.3 & 95.3 & 4.7 & 0.0 \\
    Gemini-Pro-1.5-002 & sub & 42.1 & 1.5 & 95.1 & 4.9 & 0.0 \\
    Gemini-Flash-2.0-Exp & sub & 41.8 & 1.3 & 90.0 & 7.5 & 2.5 \\
    Qwen3-Coder-30B-A3B-Instruct & sub & 38.5 & 27.5 & 92.2 & 6.9 & 0.8 \\
    GPT-4O-2024-08-06 & sub & 38.3 & 21.5 & 95.5 & 4.3 & 0.1 \\
    GPT-4-Turbo-1106 & sub & 37.4 & 26.0 & 97.5 & 2.5 & 0.0 \\
    GPT-4-Turbo-2024-04-09 & sub & 37.3 & 25.8 & 93.7 & 5.9 & 0.4 \\
    Mistral-Large & sub & 37.3 & 12.6 & 96.9 & 3.1 & 0.0 \\
    Qwen3-8B & sr & 37.0 & 12.5 & 99.4 & 0.6 & 0.0 \\
    LLama3.3-70b-Ins & sub & 36.8 & 26.1 & 70.2 & 1.0 & 28.8 \\
    Gemini-Flash-1.5-002 & sub & 36.1 & 2.1 & 85.1 & 14.9 & 0.0 \\
    GPT-4O-mini-2024-07-18 & sub & 35.5 & 21.6 & 84.9 & 14.9 & 0.2 \\
    Codestral-Latest & sub & 35.3 & 14.0 & 97.6 & 2.4 & 0.0 \\
    Qwen2.5-Coder-Ins-7B & sub & 34.1 & 8.6 & 98.7 & 1.3 & 0.0 \\
    GPT-4-0613 & sub & 33.9 & 0.9 & 94.8 & 5.2 & 0.0 \\
    Seed-Coder-8B-Instruct & sr & 33.4 & 13.2 & 96.4 & 3.6 & 0.0 \\
    Qwen2-Ins-72B & sub & 30.6 & 4.5 & 87.4 & 4.4 & 8.2 \\
    Qwen2.5-Ins-7B & sub & 30.1 & 5.4 & 98.9 & 1.1 & 0.0 \\
    DSCoder-33b-Ins & sub & 26.7 & 5.2 & 99.9 & 0.1 & 0.0 \\
    DeepSeek-Coder-V2-Lite-Instruct & sub & 25.2 & 9.5 & 96.8 & 3.2 & 0.0 \\
    GLM-4.7-Flash & sub & 24.8 & 21.2 & 95.0 & 5.0 & 0.0 \\
    Claude-3-Haiku & sub & 22.5 & 14.5 & 95.3 & 1.2 & 3.5 \\
    CodeGeeX4-All-9B & sr & 22.1 & 11.8 & 98.2 & 1.8 & 0.0 \\
    AzeroGPT-64b & sub & 22.0 & 8.0 & 70.8 & 10.7 & 18.6 \\
    DSCoder-6.7b-Ins & sub & 19.8 & 7.4 & 98.9 & 1.1 & 0.0 \\
    DSCoder-1.3b-Ins & sub & 8.6 & 5.4 & 100.0 & 0.0 & 0.0 \\
    \bottomrule
  \end{tabular}
  \caption{Full per-model form profile on LCB (part 2 of 2: remaining 46 models). Continuation of Table~\ref{tab:form-profile}.}
  \label{tab:form-profile-2}
\end{table*}

%% file: tables/06_within_model_rho.tex
\begin{table*}[t]
  \centering
  \begin{tabular}{cccc}
    \toprule
    \textbf{Form axis} & \textbf{Median $\rho$} & \textbf{Max $|\rho|$} & \textbf{\% $|\rho|{>}0.20$} \\
    \midrule
    \texttt{tok-cmt\%}  & $-0.05$ & $0.53$ & $22\%$ \\
    \texttt{block\%}    & $-0.06$ & $0.31$ & $13\%$ \\
    \texttt{inline\%}   & $-0.08$ & $0.32$ & $6\%$ \\
    \texttt{doc\%}      & $-0.01$ & $0.24$ & $4\%$ \\
    \bottomrule
  \end{tabular}
  \caption{Within-model Spearman~$\rho$ between each form metric and per-problem pass status, computed independently for each of the 88 models. Max $|\rho|$ and threshold-crossing fractions bound the share of models with non-trivial within-model variation.}
  \label{tab:within-model-rho}
\end{table*}

%% file: tables/07_decomp_random.tex
\begin{table*}[t]
  \centering
  \setlength{\tabcolsep}{4pt}
  \begin{tabular}{llcc}
    \toprule
    \textbf{Recipient} & \textbf{Source} & \textbf{Random pass@1} & \textbf{$\Delta$ vs baseline} \\
    \midrule
    \multirow{3}{*}{Gemma-4-E4B-IT} & gemini & $51.4$ & $-19.9$$^*$ \\
                                    & claude & $45.3$ & $-17.4$$^*$ \\
                                    & grok   & $42.8$ & $-21.0$$^*$ \\
    \addlinespace
    \multirow{3}{*}{Qwen3-8B}       & gemini & $23.8$ & $-29.9$$^*$ \\
                                    & claude & $35.7$ & $-10.2$$^*$ \\
                                    & grok   & $20.5$ & $-26.6$$^*$ \\
    \addlinespace
    \multirow{3}{*}{Seed-Coder-8B}  & gemini & $27.1$ & $-16.8$$^*$ \\
                                    & claude & $27.1$ & $-10.9$$^*$ \\
                                    & grok   & $23.3$ & $-15.4$$^*$ \\
    \addlinespace
    \multirow{3}{*}{CodeGeeX4-9B}   & gemini & $6.5$  & $-19.9$$^*$ \\
                                    & claude & $12.2$ & $-10.5$$^*$ \\
                                    & grok   & $6.2$  & $-16.8$$^*$ \\
    \midrule
    \multicolumn{3}{l}{mean $\Delta$ across 12 cells} & $-17.9$ \\
    \bottomrule
  \end{tabular}
  \caption{Length-matched random-text control across 12 (recipient, source) cells. The prefix is \texttt{\#}-prefixed filler matched in length to each source-written block. $\Delta$ is against the recipient's matched-prompt baseline on the same per-cell subset. $^*$ = Holm-significant against the baseline at $p_\text{holm}{<}0.001$ (paired McNemar, $m{=}12$ within this control family).}
  \label{tab:decomp-random}
\end{table*}

%% file: tables/08_difficulty.tex
\begin{table*}[t]
  \centering
  \small
  \begin{tabular}{cccccc}
    \toprule
    \textbf{Recipient} & \textbf{Source} & \textbf{Condition} & \textbf{Easy $\Delta$} & \textbf{Medium $\Delta$} & \textbf{Hard $\Delta$} \\
    \midrule
    Gemma-4-E4B-IT & gemini & source-correct & $+2.5$  & $\mathbf{+22.7}$ & $+20.3$ \\
    Gemma-4-E4B-IT & gemini & shuffled       & $-6.0$  & $\mathbf{+13.2}$ & $+11.6$ \\
    Gemma-4-E4B-IT & gemini & wrong-topic    & $\mathbf{-32.7}$ & $-19.7$ & $-17.4$ \\
    Gemma-4-E4B-IT & gemini & random-text    & $\mathbf{-27.9}$ & $-12.8$ & $-17.4$ \\
    Gemma-4-E4B-IT & claude & source-correct & $-3.8$  & $+13.3$ & $\mathbf{+17.5}$ \\
    Gemma-4-E4B-IT & claude & shuffled       & $-5.3$  & $\mathbf{+11.6}$ & $+8.3$ \\
    Gemma-4-E4B-IT & claude & wrong-topic    & $\mathbf{-30.4}$ & $-19.4$ & $-16.2$ \\
    Gemma-4-E4B-IT & claude & random-text    & $\mathbf{-23.8}$ & $-15.1$ & $-11.8$ \\
    Gemma-4-E4B-IT & grok   & source-correct & $+2.8$  & $\mathbf{+25.3}$ & $+18.9$ \\
    Gemma-4-E4B-IT & grok   & shuffled       & $+0.0$  & $\mathbf{+19.0}$ & $+11.2$ \\
    Gemma-4-E4B-IT & grok   & wrong-topic    & $\mathbf{-22.2}$ & $-14.7$ & $-15.0$ \\
    Gemma-4-E4B-IT & grok   & random-text    & $\mathbf{-33.9}$ & $-14.4$ & $-12.6$ \\
    \addlinespace
    Qwen3-8B       & gemini & source-correct & $+12.1$ & $\mathbf{+24.0}$ & $+13.8$ \\
    Qwen3-8B       & gemini & shuffled       & $+8.3$  & $\mathbf{+19.7}$ & $+13.0$ \\
    Qwen3-8B       & gemini & wrong-topic    & $\mathbf{-52.1}$ & $-24.3$ & $-14.5$ \\
    Qwen3-8B       & gemini & random-text    & $\mathbf{-43.2}$ & $-22.4$ & $-15.9$ \\
    Qwen3-8B       & claude & source-correct & $+9.1$  & $\mathbf{+20.3}$ & $+14.0$ \\
    Qwen3-8B       & claude & shuffled       & $+7.2$  & $\mathbf{+19.1}$ & $+10.1$ \\
    Qwen3-8B       & claude & wrong-topic    & $\mathbf{-39.8}$ & $-18.8$ & $-10.1$ \\
    Qwen3-8B       & claude & random-text    & $\mathbf{-11.0}$ & $-10.1$ & $-9.2$ \\
    Qwen3-8B       & grok   & source-correct & $+14.2$ & $\mathbf{+31.9}$ & $+19.4$ \\
    Qwen3-8B       & grok   & shuffled       & $+6.6$  & $\mathbf{+25.3}$ & $+17.0$ \\
    Qwen3-8B       & grok   & wrong-topic    & $\mathbf{-50.0}$ & $-21.0$ & $-10.2$ \\
    Qwen3-8B       & grok   & random-text    & $\mathbf{-46.5}$ & $-17.8$ & $-10.7$ \\
    \addlinespace
    Seed-Coder-8B  & gemini & source-correct & $+10.2$ & $\mathbf{+20.7}$ & $+10.1$ \\
    Seed-Coder-8B  & gemini & shuffled       & $+2.2$  & $\mathbf{+17.8}$ & $+6.5$ \\
    Seed-Coder-8B  & gemini & wrong-topic    & $\mathbf{-33.3}$ & $-7.2$ & $-4.3$ \\
    Seed-Coder-8B  & gemini & random-text    & $\mathbf{-35.9}$ & $-3.9$ & $-1.4$ \\
    Seed-Coder-8B  & claude & source-correct & $+11.0$ & $\mathbf{+18.0}$ & $+7.5$ \\
    Seed-Coder-8B  & claude & shuffled       & $+7.5$  & $\mathbf{+16.2}$ & $+5.3$ \\
    Seed-Coder-8B  & claude & wrong-topic    & $\mathbf{-32.0}$ & $-8.1$ & $-3.1$ \\
    Seed-Coder-8B  & claude & random-text    & $\mathbf{-23.2}$ & $-4.3$ & $-3.5$ \\
    Seed-Coder-8B  & grok   & source-correct & $+17.4$ & $\mathbf{+22.7}$ & $+11.7$ \\
    Seed-Coder-8B  & grok   & shuffled       & $+4.7$  & $\mathbf{+16.7}$ & $+7.3$ \\
    Seed-Coder-8B  & grok   & wrong-topic    & $\mathbf{-26.9}$ & $-11.5$ & $-3.9$ \\
    Seed-Coder-8B  & grok   & random-text    & $\mathbf{-35.1}$ & $-4.6$ & $-3.4$ \\
    \addlinespace
    CodeGeeX4-9B   & gemini & source-correct & $+27.3$ & $\mathbf{+33.2}$ & $+6.5$ \\
    CodeGeeX4-9B   & gemini & shuffled       & $+17.8$ & $\mathbf{+24.7}$ & $+9.4$ \\
    CodeGeeX4-9B   & gemini & wrong-topic    & $\mathbf{-38.7}$ & $-3.6$ & $-2.2$ \\
    CodeGeeX4-9B   & gemini & random-text    & $\mathbf{-41.0}$ & $-6.2$ & $-2.2$ \\
    CodeGeeX4-9B   & claude & source-correct & $+19.4$ & $\mathbf{+24.6}$ & $+8.8$ \\
    CodeGeeX4-9B   & claude & shuffled       & $+17.9$ & $\mathbf{+21.7}$ & $+5.7$ \\
    CodeGeeX4-9B   & claude & wrong-topic    & $\mathbf{-39.2}$ & $-4.9$ & $-1.3$ \\
    CodeGeeX4-9B   & claude & random-text    & $\mathbf{-23.5}$ & $-4.6$ & $-1.3$ \\
    CodeGeeX4-9B   & grok   & source-correct & $+25.9$ & $\mathbf{+29.9}$ & $+10.2$ \\
    CodeGeeX4-9B   & grok   & shuffled       & $+6.6$  & $\mathbf{+23.0}$ & $+5.8$ \\
    CodeGeeX4-9B   & grok   & wrong-topic    & $\mathbf{-34.8}$ & $-6.3$ & $-0.5$ \\
    CodeGeeX4-9B   & grok   & random-text    & $\mathbf{-40.2}$ & $-4.9$ & $-1.0$ \\
    \bottomrule
  \end{tabular}
  \caption{$\Delta$ pass@1 (\%) against the recipient's matched-prompt baseline on the source-pass subset, stratified by LCB difficulty band. Recipients sorted by base pass@1 descending. \textbf{Bold} = largest-magnitude band per row.}
  \label{tab:difficulty}
\end{table*}

%% file: tables/09_per_model_audit.tex
\begin{table*}[t]
  \centering
  \setlength{\tabcolsep}{6pt}
  \begin{tabular}{lccc}
  \toprule
  \textbf{Model (base \%)} & \textbf{Best (variant, $\Delta$\%)} & \textbf{Worst (variant, $\Delta$\%)} & \textbf{\#sig} \\
  \midrule
  Grok-4.1-Fast (85.1)         & inline $-0.3$           & \textbf{WHY $-4.1$}            & 1 \\
  Gemini-3.1-Flash-Lite (72.5) & inline $+0.2$           & \textbf{none $-6.9$}           & 3 \\
  GPT-OSS-120B (69.1)          & \textbf{mix $+7.2$}     & \textbf{inline $-13.9$}        & 5 \\
  GPT-5.1-Codex-Mini (64.6)    & \textbf{WHAT $+4.9$}    & \textbf{top $-13.5$}           & 6 \\
  DeepSeek-V3.2 (64.3)         & \textbf{top $+6.2$}     & \textbf{none $-6.1$}           & 4 \\
  Qwen3-235B (59.8)            & \textbf{WHY $+5.7$}     & none $-1.5$                    & 1 \\
  GPT-5.4-Nano (59.6)          & \textbf{dense $+5.8$}   & \textbf{none $-13.6$}          & 3 \\
  MiMo-V2-Flash (55.9)         & \textbf{WHY $+7.2$}     & \textbf{top $-9.0$}            & 3 \\
  Gemma-4-E4B-IT (53.7)        & top $+0.5$              & \textbf{none $-14.5$}          & 5 \\
  Llama-4-Maverick (45.3)      & inline $+0.3$           & \textbf{WHAT $-12.7$}          & 3 \\
  Mistral-Small (44.2)         & \textbf{dense $+4.4$}   & none $-1.7$                    & 1 \\
  Qwen3-8B (37.0)              & \textbf{dense $+4.4$}   & none $-1.6$                    & 1 \\
  Seed-Coder-8B (33.4)         & inline $+2.8$           & dense $-3.9$                   & 0 \\
  CodeGeeX4-9B (22.1)          & inline $-0.6$           & \textbf{WHY $-5.0$}            & 2 \\
  \bottomrule
  \end{tabular}
  \caption{Per-model dramatic-cell audit, sorted by \textbf{base} pass@1 descending. \#sig: Holm-significant cells per model (global Holm over $m{=}126$, Table~\ref{tab:grid-14x10}). \textbf{Bold} $\Delta$ = Holm-significant.}
  \label{tab:per-model-audit}
\end{table*}

%% file: tables/10_variant_compliance.tex
\begin{table*}[t]
  \centering
  \begin{tabular}{lrr}
    \toprule
    \textbf{Variant} & \textbf{Mean tokens} & \textbf{Mean tok-cmt\%} \\
    \midrule
    base               & 463 & 22.7 \\
    \textbf{none}      & 368 & \textbf{6.3} \\
    rare               & 411 & 14.9 \\
    dense              & 740 & 55.4 \\
    WHAT               & 566 & 49.6 \\
    WHY                & 658 & 54.6 \\
    mix                & 685 & 55.6 \\
    gp-doc             & 594 & 45.5 \\
    top                & 655 & 57.3 \\
    inline             & 604 & 56.7 \\
    \bottomrule
  \end{tabular}
  \caption{Per-variant mean output code tokens and tok-cmt\%, averaged across the 14 models in the self-elicit grid (column means of Table~\ref{tab:per-cell-grid-form}). Models reliably comply with surface form instructions, but compliance does not track pass@1 effect (Section~\ref{sec:selfelicit-results}).}
  \label{tab:variant-compliance}
\end{table*}

%% file: tables/11_per_cell_grid_form.tex
\begin{table*}[t]
  \centering
  \scriptsize
  \setlength{\tabcolsep}{4pt}
  \begin{tabular}{lcccccccccc}
    \toprule
    \textbf{Model} & \textbf{base} & \textbf{none} & \textbf{rare} & \textbf{dense} & \textbf{WHAT} & \textbf{WHY} & \textbf{mix} & \textbf{gp-doc} & \textbf{top} & \textbf{inline} \\
    \midrule
    Grok-4.1-Fast         & 277/1.4    & 267/0.3    & 269/0.6    & 479/46.6   & 399/43.5   & 510/57.8   & 524/57.2   & 295/16.9   & 456/51.5   & 383/36.3   \\
    Gemini-3.1-Flash-Lite & 440/39.5   & 235/0.1    & 255/18.8   & 541/59.9   & 441/51.6   & 556/61.3   & 575/62.7   & 458/52.3   & 629/66.0   & 581/61.6   \\
    GPT-OSS-120B          & 224/15.7   & 183/1.4    & 195/6.5    & 372/52.3   & 352/38.7   & 381/44.7   & 416/48.4   & 400/48.0   & 437/64.7   & 388/54.6   \\
    GPT-5.1-Codex-Mini    & 154/0.6    & 130/0.1    & 138/0.2    & 286/55.8   & 383/53.7   & 353/50.6   & 393/53.5   & 249/31.4   & 259/44.5   & 310/52.9   \\
    DeepSeek-V3.2         & 603/22.6   & 269/0.1    & 239/5.9    & 929/56.0   & 671/44.2   & 803/49.6   & 826/53.1   & 796/49.0   & 1094/77.6  & 1022/72.8  \\
    Qwen3-235B-A22B       & 495/37.7   & 429/26.6   & 384/27.2   & 811/64.1   & 821/55.4   & 637/48.8   & 923/57.6   & 747/52.1   & 1057/56.8  & 778/65.3   \\
    GPT-5.4-Nano          & 679/39.4   & 486/9.4    & 613/29.9   & 843/64.5   & 821/56.8   & 862/60.4   & 916/62.7   & 802/51.9   & 776/64.5   & 839/63.9   \\
    MiMo-V2-Flash         & 1119/36.3  & 1628/24.6  & 1864/28.2  & 2722/66.7  & 1168/58.0  & 1582/69.1  & 1309/65.8  & 1400/59.2  & 992/59.3   & 767/61.7   \\
    Gemma-4-E4B-IT        & 1067/58.6  & 437/7.3    & 668/42.0   & 1222/73.2  & 1003/68.2  & 1181/72.2  & 1218/72.5  & 986/65.2   & 1246/71.7  & 1168/71.9  \\
    Llama-4-Maverick      & 239/11.3   & 193/0.6    & 222/12.9   & 365/46.8   & 316/38.4   & 318/34.4   & 286/26.0   & 278/28.2   & 228/15.3   & 405/52.3   \\
    Mistral-Small-2603    & 365/16.9   & 207/0.3    & 210/2.6    & 657/60.4   & 574/55.3   & 772/65.9   & 979/65.2   & 704/56.0   & 953/76.6   & 740/66.1   \\
    Qwen3-8B              & 399/12.5   & 347/6.4    & 309/9.9    & 642/51.1   & 465/48.1   & 669/57.3   & 659/60.0   & 665/53.5   & 474/50.7   & 584/49.8   \\
    Seed-Coder-8B         & 207/13.2   & 170/3.5    & 190/8.9    & 225/37.2   & 224/37.2   & 294/46.7   & 280/46.2   & 259/33.0   & 298/52.9   & 233/39.0   \\
    CodeGeeX4-9B          & 210/11.8   & 176/7.1    & 202/14.7   & 265/40.6   & 279/44.7   & 296/46.0   & 282/47.4   & 276/40.0   & 270/50.4   & 265/45.8   \\
    \bottomrule
  \end{tabular}
  \caption{Per-cell mean output code tokens and tok-cmt\% across the 14$\times$10 self-elicit grid (Section~\ref{sec:self-elicit}), sorted by \textbf{base} pass@1 descending. Each cell shows $\text{tok} / \text{cmt\%}$: mean output code tokens under \texttt{cl100k\_base}, and the mean fraction of those tokens that fall inside a comment.}
  \label{tab:per-cell-grid-form}
\end{table*}